\documentclass{article}

\usepackage[preprint]{Learn_scheduling}
\usepackage[utf8]{inputenc} 
\usepackage[T1]{fontenc}    
\usepackage{hyperref}       
\usepackage{url}            
\usepackage{booktabs}       
\usepackage{amsfonts}       
\usepackage{nicefrac}       
\usepackage{microtype}      
\usepackage{xcolor}         

\usepackage{amsmath}
\usepackage{mathtools}
\usepackage{multirow, booktabs}
\usepackage{pdfpages}
\usepackage{enumitem}
\usepackage{xcolor}
\usepackage{pgfplots}
\usepackage[titlenumbered, linesnumbered,lined]{algorithm2e}
\RestyleAlgo{ruled}
\usepackage{algorithmic}
\usepackage{float}  
\usepackage{lipsum}
\usepackage{caption}
\usepackage{subcaption}
\usepackage{verbatim}
\usepackage{wrapfig}

\title{Learning Interpretable Scheduling Algorithms for Data
Processing Clusters}

\author{%
  Zhibo Hu \\
  The University of New South Wales \\
  CSIRO Data61 \\
  Australia \\
  \texttt{zhibo.hu@student.unsw.edu.au} \\
  \And
  Chen Wang \\
  CSIRO Data61 \\
  The University of New South Wales \\
  Australia \\
  \texttt{chen.wang@data61.csiro.au} \\
  \And
  Helen Hye-Young Paik \\
  The University of New South Wales \\
  Australia \\
  \texttt{h.paik@unsw.edu.au} \\
  \AND
  Yanfeng Shu \\
  CSIRO Data61 \\
  Australia \\
  \texttt{yanfeng.shu@data61.csiro.au} \\
  \And
  Liming Zhu \\
  CSIRO Data61 \\
  The University of New South Wales \\
  Australia \\
  \texttt{liming.zhu@data61.csiro.au} \\
}

\begin{document}

\maketitle

\begin{abstract}
  Workloads in data processing clusters are often represented in the form of DAG (Directed Acyclic Graph) jobs. Scheduling DAG jobs is challenging. Simple heuristic scheduling algorithms are often adopted in practice in production data centres. There is much room for scheduling performance optimisation for cost saving. Recently, reinforcement learning approaches (like decima) have been attempted to optimise DAG job scheduling and demonstrate clear performance gain in comparison to traditional algorithms. However, reinforcement learning (RL) approaches face their own problems in real-world deployment. In particular, their black-box decision making processes and generalizability in unseen workloads may add a non-trivial burden to the cluster administrators. Moreover, adapting RL models on unseen workloads often requires significant amount of training data, which leaves edge cases run in a sub-optimal mode. To fill the gap, we propose a new method to distill a simple scheduling policy based on observations of the behaviours of a complex deep learning model. The simple model not only provides interpretability of scheduling decisions, but also adaptive to edge cases easily through tuning. We show that our method achieves high fidelity to the decisions made by deep learning models and outperforms these models when additional heuristics are taken into account. 
\end{abstract}

\section{Introduction}
In many data processing clusters, data analytics software stack 
often provides a programming model for developers to express data parallelism so that 
job scheduling algorithms can optimise the resource allocation (e.g., SparkSQL~\citep{armbrust2015spark} on Spark~\citep{zaharia2012resilient}, Dryad~\citep{isard2007dryad} and TEZ~\citep{tez}). 

Commonly, user applications are broken into a series of ``compute'' jobs where each job  contains a number of interdependent tasks. DAGs, or direct acyclic graphs are used to represent the input-output dependency among these tasks. Each vertex is a node that contains a number of homogeneous tasks and the edge between two nodes shows one  requires the other's output.    

\cite{grandl2016graphene} have shown that in production clusters, jobs often have large and complex DAGs. Improving the efficiency of a scheduler could save operation costs, even a small amount of improvement in utilization can save millions of
dollars at scale \citep{barroso2013datacenter}.
However, how to schedule these jobs for efficient resource utilization is still challenging and conventional approaches rely on heuristic algorithms.  For instance, Dryad~\citep{isard2007dryad} adopts a simple greedy algorithm that allocates a vertex that is ready to run to an available job executor under the assumption that the job is the only workload in the cluster. Capacity Schedulers~\citep{capacityschedule} simply picks a DAG that is ready to run to fit the available executor resources. GRAPHENE \citep{grandl2016g} uses a two-level scheduling strategy with the intra-DAG scheduler. 
It divides a DAG into sub-graphs which are then ranked according to a heuristic scoring method that considers task arrival times, remaining work, locality and fairness constraints.

In recent years, machine learning based methods have been increasingly used to optimise complex operations. In database systems, SageDB\citep{kraska2021sagedb}) and Neo~\citep{marcus2019neo}) learn optimal data access methods and database query plans and \citet{krishnan2018learning} learns to optimise join queries. In  networks, RL-cache~\citep{kirilin2019rl} learns caching admission policies for content delivery networks (CDNs), while deep reinforcement learning is used for network planing~\citep{zhu2021network}. Very recently, reinforcement learning methods have discovered new sorting algorithms that outperform known human benchmarks~\citep{Mankowitz2023}.

Similarly, deep learning methods~\citep{peng2021dl2, mao2019learning} have been exploited for the resource scheduling problem. Notably, \citet{mao2019learning} develops Decima scheduler which uses 

Graph Neural Networks (GNN) to learn representations of DAG jobs so that a Deep Reinforcement Learning (DRL) framework can obtain a scheduling policy for the complex jobs that are difficult to charaterise using traditional methods. These deep learning based methods have shown superior performance compared to traditional methods. 

However, there is a significant gap to be filled for Reinforcement learning (RL) based schedulers to be deployed in practice. The main problem is that the learned models are complex and the scheduling decisions are difficult to understand. Further, it is challenging to patch a model when problems occur on edge cases. 
We observe heuristic algorithms outperform the RL model on some individual cases, but there is no easy solution to tune an RL model to accommodate these cases. It is difficult to know why the RL model fails on these cases, therefore unrealistic to provide a certain number of similar cases for training. Directly tuning a complex model may incur problems  such as "forgetting of pre-trained capabilities" \citep{wolczyk2024fine}.

In this paper, we intend to address these problems by leveraging interpretable models derived from complex RL models. We give a method to distill interpretable, or decision tree based schedulers from the RL scheduler behaviours and then using the interpretable schedulers for scheduling and patching for edge cases. As RL can be viewed as doing supervised learning on the “good data” (optimized data) \citep{eysenbach2020reinforcement}, an interpretable scheduler is a supervised model trained on RL's decisions made on those data. Importantly, we show that multiple interpretable models can be derived from the same RL model and a dynamically selection algorithm for these models leads to the good scheduling adaptability to edge cases.  
\begin{figure}[!htb]
    \centering
    \includegraphics[width=.8\linewidth]{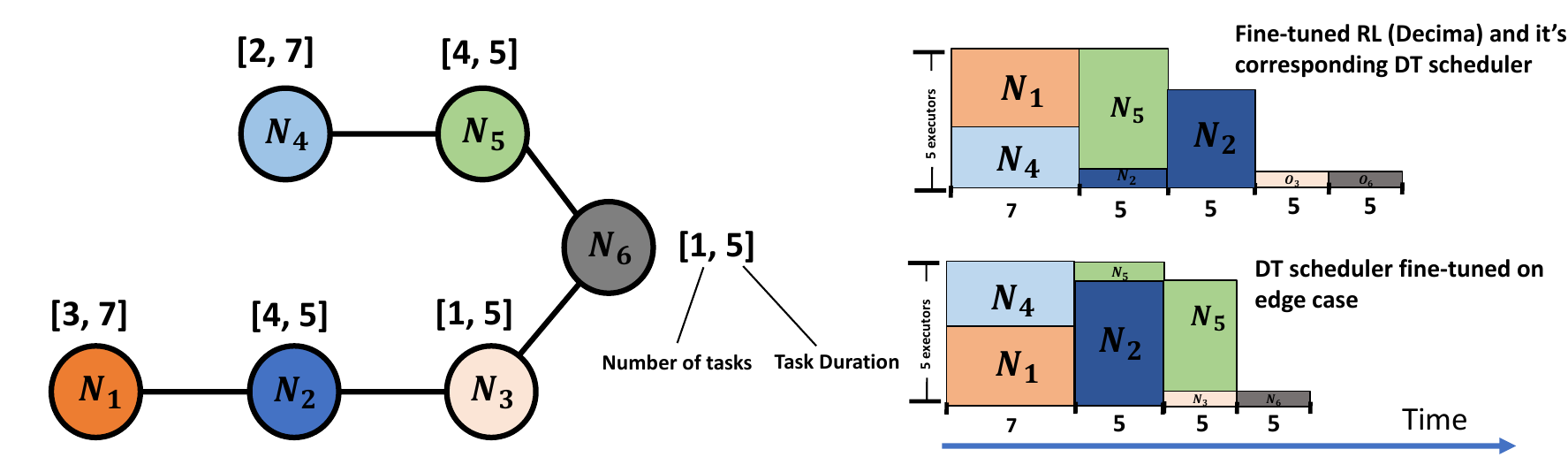}
    \caption{Example on the scheduling quality of different schedulers: Decima and its corresponding decision tree scheduler have the same sub-optimal scheduling trace (top right), while the decision tree scheduler fine-tuned on additional traces such as edge cases has the improved scheduling trace for this job DAG (bottom right). } 
\label{fig:motivation_example}
\end{figure}

A decision tree is sensitive to data distribution change~\citep{liao2024invariant}, however, this property also makes a decision tree adaptive when adding edge cases into training. Fig.~\ref{fig:motivation_example} shows an example of scheduling decisions made by three models. The fine-tuned decision tree scheduler produces the best performance for the workload.

Specifically, this paper makes the following contributions: (1) We propose a DNN model interpretability method that utilises human knowledge to derive a decision tree model that clearly shows how the DNN model makes the performance gain; (2) We design a scheduler for data processing clusters that outperforms the state-of-the-art scheduling algorithm on job completion time; (3) We provide a solution to the problem where the RL model cannot be efficiently fine-tuned to accommodate edge cases. By transferring the RL setting to a supervised learning on optimized data, we are able to select and tune the interpretable models efficiently. 

\section{Related Work} \label{sec:lit}
\paragraph{Data parallel cluster scheduling.}
Many DAG job scheduling research either focuses on the theoretical aspect~\citep{shmoys1994improved,chekuri2004multi}, or adopts simple greedy algorithms that schedules a task once its dependencies are solved and the necessary resources are available~\citep{hindman2011mesos, isard2007dryad, yarnfairschedule}. Decima~\citep{mao2019learning} uses GNN to learn embeddings of job DAGs and feed them to a reinforcement learning (RL) framework to obtain a scheduling policy that fully leverages the dependency graph information. Decima shows a clear performance advantage over greedy and fair scheduling algorithms. 
LSched~\citep{sabek2022lsched} offers efficient inter-query and intra-query scheduling for dynamic analytical workloads. Despite this, RL-based schedulers like Decima and LShed are computationally expensive, highlighting the need for more efficient ML schedulers. Recently, \citet{jeon2022neural} propose a ML scheduler that samples node priorities using a one-shot neural network encoder and then utilizes list scheduling to create the final schedules. This scheduler runs much faster than existing ML baselines because the one-shot encoder can efficiently sample priorities in parallel. 
\paragraph{Model interpretation.}
There are two main approaches developed to explain and understand deep learning models: post-hoc explanations and built-in explainable models. Post-hoc explanation techniques seek insight into a trained DNN model to find links to prediction results~\citep{8466590}, while build-in explainable models intentionally guide model decisions to focus certain part of the input or relate to certain ``prototype'' in reasoning to achieve end-to-end interpretability~\citep{chen2019looks}. Post-hoc approaches are criticised for not faithfully reflecting what the original models do~\citep{rudin2019stop}, undermining user trust~\cite{chu2020visual}. \citet{rudin2019stop} suggests that high-stakes applications should use built-in explainable models when possible. However, a build-in explainable model, such as decision trees and rule-based models, can only deal with problems with relatively low complexity. Recent approaches show certain success in injecting reasoning processes into complex models. \citet{li2018deep} introduces a prototype layer in a DNN model to find parts of training images that represent a specific class. Another approach, \emph{mimic learning}~\citep{che2016interpretable}, distills soft labels from deep networks to learn interpretable models such as Gradient Boosting Trees. Similarly, \citet{meng2020interpreting} uses a teacher-student training approach to obtain an interpretable model based on a DNN model for networking systems. Compared to networking systems, a DAG job scheduling system deals with more complex inputs with dependency and a larger space for resource mapping. Moreover, none of these interpretability work focus on correcting problems in original DNN models. 

\section{Decision Tree Scheduler}\label{sec:methods}

\begin{figure}[!htb]
    \centering
    \includegraphics[width=0.5\linewidth]{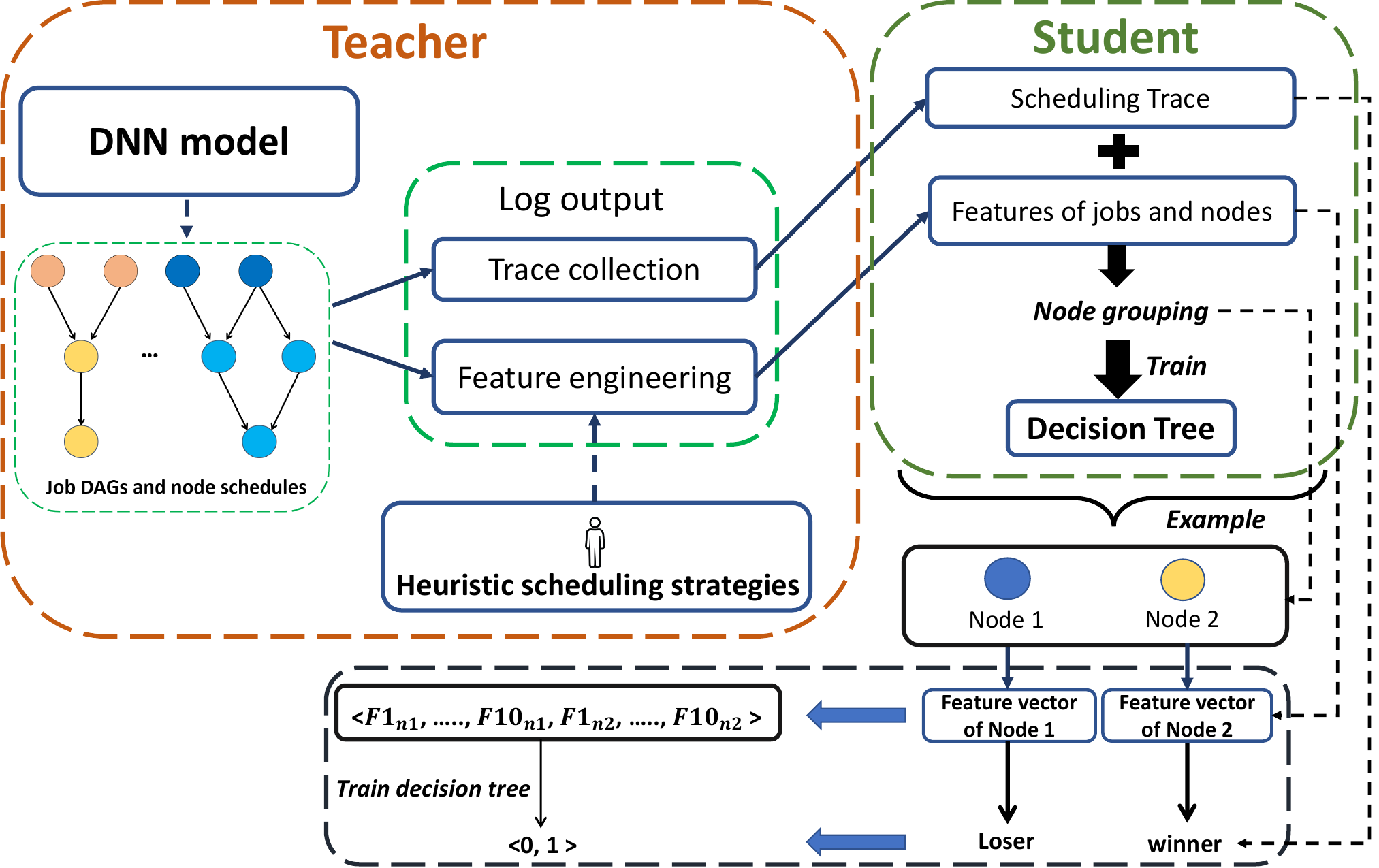}
    \caption{The process of extracting scheduling strategies from a DNN-based scheduler and constructing decision trees. In the decision tree construction example, nodes in the trace are assigned to groups and node features are constructed based on human knowledge. The winning nodes (nodes selected to schedule) are given by the DNN model's scheduling decisions in each stage. 
}
\label{fig:Distill_strategies_details}
\end{figure}

As shown in Fig.~\ref{fig:Distill_strategies_details}, we use a teacher-student framework to train a simple decision tree scheduler. The DNN-based scheduler is the teacher and the decision tree scheduler is the student. The student model learns from the scheduling trace of the teacher model with an objective of matching the scheduling decisions. We construct the feature set of the decision tree using knowledge accumulated in existing heuristic DAG scheduling literature to achieve interpretability.

After obtaining the decision tree, we are able to improve it further by integrating the interpretable tree with a fair scheduling strategy and further tune the tree to adapt to jobs scheduled in a sub-optimal way. The resulting new scheduler shows surprising performance gain against the DNN-based scheduler. 

In this section, we first describe the process of collecting the prediction sequence/logs from the DNN model, and then detail the process of constructing a decision tree and how to tune it for edge cases.

\subsection{DNN scheduling traces}\label{sec:dnntrace}
As mentioned earlier, a typical DNN scheduler for DAG jobs uses a graph neural network to obtain representations of the DAGs. Then a policy network is considered to obtain a schedule that minimizes a given metric such as average JCT (job completion time). A job is represented as a set of inter-dependent nodes. Each node contains a number of homogeneous tasks that run on data partitions. The learned model makes scheduling decision based on scheduling stages. A scheduling stage is triggered by a scheduling event like an executor becoming idle, an arrival of a new job, or a resolution of a dependency. 

Using Decima as an example, a scheduling decision in the trace can be described as ``node $v$ of job $i$ being selected to run on $l_i$ executors''. $l_i$ is also referred to as the maximum parallelism for job $i$. 
At each scheduling stage, we record the following information: $(i, v, l_i, I \setminus \{i\}, R)$, in which $I$ is the sequence of jobs ready to schedule at this stage. $I$ is maintained by the \emph{job manager} of the scheduling system. $R$ is the executor status in the cluster. The feature set of job $i$, denoted by $x_i$ can be retrieved from the \emph{job manager}. $x_i$ contains the DAG with each node in the DAG associated with attributes like the number of remaining tasks of the node and expected execution time of each task. It also contains the information of executors allocated to the job. A \emph{resource manager} monitors the status of executors in the cluster and handles the dispatch requests from the job manager. 

A DNN scheduler computes the node embedding and DAG embedding based on the raw features. The scheduling policy is trained to select a node to run on a number of executors to minimise the average number of jobs in the system, hence the average JCT of jobs. How a node is selected is not transparent in the process as the mapping between raw features to embedding is non-linear and it is difficult to attribute the decision to job or node level features through dynamic embedding. To address the problem, we first  construct a set of features used in traditional scheduling heuristics, and then build an interpretable model to mimic the DNN scheduler decisions based on these features. The question we are asking is whether the DNN scheduling decisions can be explained using commonly used features.

The feature set of a node includes two parts: raw features and composite features. 
The raw features include the following: (1) the current number of executors assigned to the job of the node(denoted by F1); (2) the flag indicating whether there are executors just finish executing a task of the job (F2); (3) the number of idle executors (F3);(4) the remaining workload of this node (F4); (5) the number of remaining tasks of the node (F5).

The composite features aggregate information from multiple nodes starting from a node in the DAG. These features include the following:(1) the total number of tasks at the nodes along the longest path (or a critical path), starting from the current node on the DAG including the current node. We apply \emph{Topological Sorting} to find the longest path (F6); (2) the total remaining workload of the nodes on the longest path starting from the current node (F7); (3) the total  number of remaining tasks of the job of the node (F8); (4) the total remaining workload of the job to which the node belongs (F9); (5) whether the node scheduled in the previous stage belongs to the same job of the current node (F10). This is to exploit the locality of resource allocation as nodes of the same job often uses the same dataset.

These features are commonly used in heuristic scheduling algorithms to characterise the size of jobs. This is a key step to understand the DNN scheduling strategy, which can be seen as expressing the DNN decisions with the known vocabulary of heuristic scheduling. The values of these features are normalised against standard computing capacity in the DNN model training and our experiments. We will show in section~\ref{sec:res_explain} that this helps reveal the scheduling strategy of the DNN model in a human understandable way. 

\subsection{Trace mimicking strategies}
With the trace, $\{(i, v, l_i, I \setminus \{i\}, R)\}$ collected from each scheduling stage, we can develop a model to mimic the scheduling decisions in the trace.

The knowledge distillation method~\citep{hinton2015distilling} uses a teacher-student training framework to learn a simple model based on the numerical information in the last few layers in the DNN model. However, there are two challenges with this approach. A well-designed DNN scheduler such as Decima~\citep{mao2019learning} often introduces certain randomness in the selection of top-$k$ candidate nodes with the highest softmax scores. The purpose is to improve the robustness of scheduling decisions. Experiments indeed show significant scheduling performance improvement with randomness. However, the randomness introduces noise in the student model training, which makes it difficult to interpret the model decisions in a consistent manner.

We therefore introduce a method that decomposes the trace mimicking task into two parts: first, we assign nodes of jobs to fixed-sized groups (node grouping) and build a decision tree to learn the node selection strategy within these groups; second, we use the decision tree as a comparator and apply it to nodes of all jobs in the queue to select the next node to run in a scheduling stage. By doing so, we are able to achieve the interpretability goal and the scheduling mimicking goal simultaneously. For the interpretability, the decision tree explains why a node is selected in a given group. For scheduling mimicking, the optimisation process minimises the difference of node selection between the DNN-based scheduler and the student scheduler in all scheduling stages.

\SetKwComment{Comment}{//}{} 

\begin{algorithm}[t]
  \SetKwInOut{Input}{Input}
  \SetKwInOut{Output}{Output}
  \begin{footnotesize} 
  
  \Input {Scheduling trace containing the following: $\mathcal{J}$ -- A list of schedulable jobs; $\mathcal{V}$ -- the schedulable nodes; $\mathcal{Y}_V$ -- the DNN model output about the selection probabilities of schedulable nodes; $i_{prev}$ -- job scheduled previously; 
  $|g|$ -- the group size.  
  } 
  \Output{ the decision tree for node selection  }

  \BlankLine
  \Comment{initialise the training set}
  $L \leftarrow \{\} $ \\
    \For{each scheduling stage in the trace}{
        Find the node with the highest selection probability from $\mathcal{Y}_V$, denoted by $v_w$ and its corresponding job by $i_w$. \\
        Obtain the composite features for $v_w$ \\
        \For{$i \in \mathcal{J} \setminus \{i_w\}$}{
        \Comment{set the locality flag}
        \eIf {$i_{prev} == i$} {
            set scheduling locality of $v \in i$ as true         
        } {
            set scheduling locality of $v \in i$ as false
            }
        }
        \Comment{add nodes to the group and add the group to the training set}
        \For{each combination $\{v_k | v_k \in \mathcal{V} \}|_{k=1}^{|g|-1}$}{
        obtain composite node features for $v_k$  \\
        add $((v_1,...,v_{|g|-1}, v_w), v_w)$ to $L$ 
        }
    }
    fit a decision tree classifier $f$ on $L$ where $v_w$ is used as the label for a record in $L$ \\ 
    \Return{the decision tree $f$}.
  \end{footnotesize}
  \caption{Decision tree construction}
  \label{alg:tree_construction}

\end{algorithm}\DecMargin{1em}

\subsection{Decision tree as a node comparator}

\paragraph{Decision Tree Construction} Algorithm~\ref{alg:tree_construction} details how the decision tree is learned. The key idea is to form groups of contrastive nodes in training. The groups are formed in fixed sizes, denoted by $|g|$. Each group contains one node with the highest probability to be selected by the DNN model for scheduling and $|g|-1$ nodes that are not selected for scheduling. We call the former ``winning node'' and the latter ``losing nodes''. By doing so, we reduce the complexity of predicting the node to schedule next so that an interpretable model can learn the difference between the ``winning'' and ``losing'' class. To ensure the interpretability of the simple model, we use a small group size of 2 (pair) or 3 (triplet). The group size is configurable for different application needs. To reduce the total number of groups to compare during scheduling, we adopt a pruning strategy as follows: when a node $v$ fails to be the ``winning node'' in $m$ groups, it is removed from the ready node and the groups are re-assigned. With this strategy, the running time of the scheduler will not exponentially increase with the number of ready nodes as most of nodes are excluded during the group assignment phase. 

The feature vectors (F1 to F10) of all the nodes in each group are concatenated together. Then the decision tree is trained by fitting the concatenated feature vector of each group with the winner's label in that group.
In addition, the depth of the learned decision tree is also adjustable for obtaining different views of the DNN schduler behaviours. We keep a given number of trees that best fit the DNN model decisions in a pool for edge case handling to be described in Section~\ref{sec:edge_case}.

\paragraph{Scheduler Construction} After training such decision trees, we integrating the best-fit interpretable tree with a fair scheduling strategy and we call the resulting new scheduler as \emph{decision tree scheduler}. 
In a data processing cluster, the job manager and resource manager feed job DAGs and executor status to the scheduler. For each scheduling stage, the scheduler assigns nodes with resolved dependency 
into groups. Each group contains a node selected to schedule and $|g|-1$ nodes that are not selected to schedule in this stage. 
The learned decision tree predicts the winning node in each group. A ``winning counter'' is maintained to track the number a node is selected in different groups. The process repeats until each node has been compared with other nodes in groups. The final winner is decided by the ``winning counter''. The node that receives the highest winning count is scheduled to run on available executors. When multiple nodes have the same number of ``winning counts'', depending on the optimisation objective, we may choose the one based on the rank of a specific feature, e.g., when the optimisation objective is to minimise the JCT, we select a node with the least work load to schedule among nodes with equal winning count. In this process, the decision tree plays a role of node comparator that decides which node to schedule. 

\subsection{Allocation of executors}
In addition to predict which node to schedule, a job scheduler for data processing cluster also needs to allocate a specific number of executors to the node. For a Spark job, this is often done through setting the upper limit of the number of executors allocated to a node. The executor allocation can be learned based on the DNN scheduler trace in a similar way as the node selection. It can also incorporate a different executor allocation strategy. We adopt a fair scheduling strategy~\cite{ghodsi2011dominant}, which splits the total executor number among the current remaining jobs in the system equally. The design choice is based on our observation from the experiment. We find that for a DNN scheduler, simultaneously learning node selection and executor allocation does not lead to a superior scheduling performance. It is likely due to a significantly larger search space to find an optimal combination of node and executor number. By replacing the executor allocation with a fair scheduling strategy, we achieve a significantly better performance than the original DNN scheduler. Such a surprising result indicates that decomposing a complex DNN scheduler into an interpretable scheduler and a well-understood resource allocation strategy not only helps the deployment of the scheduler in practice, but also realises the potential of improving the whole scheduling performance.

\subsection{Multiple decision trees and edge case tuning}\label{sec:edge_case}
As mentioned, we keep multiple decision trees that fit the DNN scheduler decisions the best. Each tree contains a slightly different explanation to the DNN scheduler's decision making process. These interpretable trees enable intervention to scheduling strategies when sub-optimal cases are identified. A sub-optimal case indicates that the tree does not provide a satisfactory explanation for the scheduling decision for a specific job. We show a better explanation can be found through tuning the tree with the job. 

The tuning process is simple: we first use a heuristic scheduler to schedule jobs in the trace containing the specific job and obtain an improved scheduling decisions; we then feed the trace to tune the decision trees in the pool and evaluate their performance on the specific job; finally we select the tree that achieves the best performance for the job and use the tree for subsequent scheduling. 
Our experiment results show that the tuned tree often accommodates the specific case by growing branches without affecting the overall performance of other jobs.

\section{Experiments}

In this section, we evaluate our decision tree scheduler by addressing the following questions: (1) How interpretable are the decision trees learned from DNN models? (2) How does the scheduler perform compared to Decima and heuristic scheduling strategies? and (3) How well does our scheduler adapt to edge cases? For our evaluation, we use TPC-H~\citep{TPC-H} and a production workload from Alibaba job-trace \citep{Alibaba-trace-2018} as the benchmark workloads. 
We also use a faithful simulator~\citep{mao2019learning} as the Spark cluster testing environment.

\subsection{Interpretability of Decision Trees Learned from DNN}\label{sec:res_explain}

\noindent\textbf{Fidelity to DNN node selection decisions.} We first evaluate the fidelity of the decision tree in node selection to the original DNN model. To ensure that the jobs in the test set are not seen by the decision tree scheduler, we separate jobs into two non-overlapping partitions. We run Decima on these two partitions and from the logs, we obtain a training set and a test set for the decision tree scheduler. Details of the datasets generated can be found in Appendix~\ref{sec:appendix_interpretability_dataset_details}.

Table~\ref{tab:fidelity} shows the fidelity of the decision tree's ``winning node'' choices to 
those selected by Decima. 
We measure fidelity by accuracy, and compare the fidelity in node selection within a group and across the entire 
test trace. As shown in the table, the accuracy is high for TPC-H jobs. 
However, accuracy is lower with a group size of 3, primarily because 
there are more candidates to select from as 
the group size increases. 
This demonstrates that our decision tree method can learn Decima's scheduling decision strategy with high fidelity, providing 
a largely faithful interpretation of the decisions made by the blackbox model.

\begin{table}[htb]
    \caption{The accuracy of selecting ``winning nodes'' on TPC-H.}
    \begin{center}
    \footnotesize
    \begin{tabular}{@{}|c|cc|@{}}
     \hline
     |g| & within group & across test trace \\
     \hline
     2 & 0.943 & 0.918 \\
     3 & 0.903 & 0.899 \\
     \hline
    \end{tabular}
    \end{center}
    \label{tab:fidelity}
\end{table}

\begin{figure}[htb]
    \centering
    \includegraphics[width=.9\linewidth]{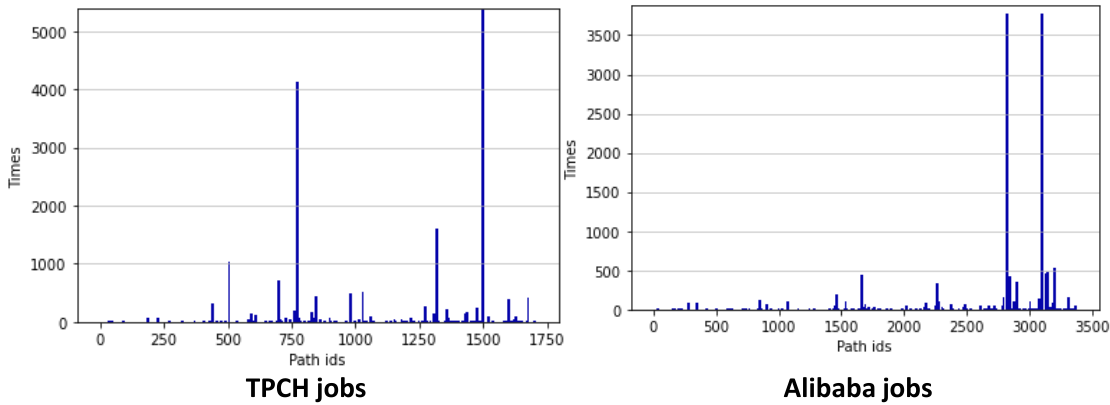}
    \caption{Decision tree trajectory distribution of TPC-H jobs and Alibaba jobs. X-axis represents the path IDs and y-axis is the number of times a decision path is taken.
}
\label{fig:Leave_distribution}
\end{figure}

\begin{figure}[htb]
  \centering
  \begin{subfigure}[b]{0.49\linewidth}
     \centering
     {\includegraphics[width=\linewidth]{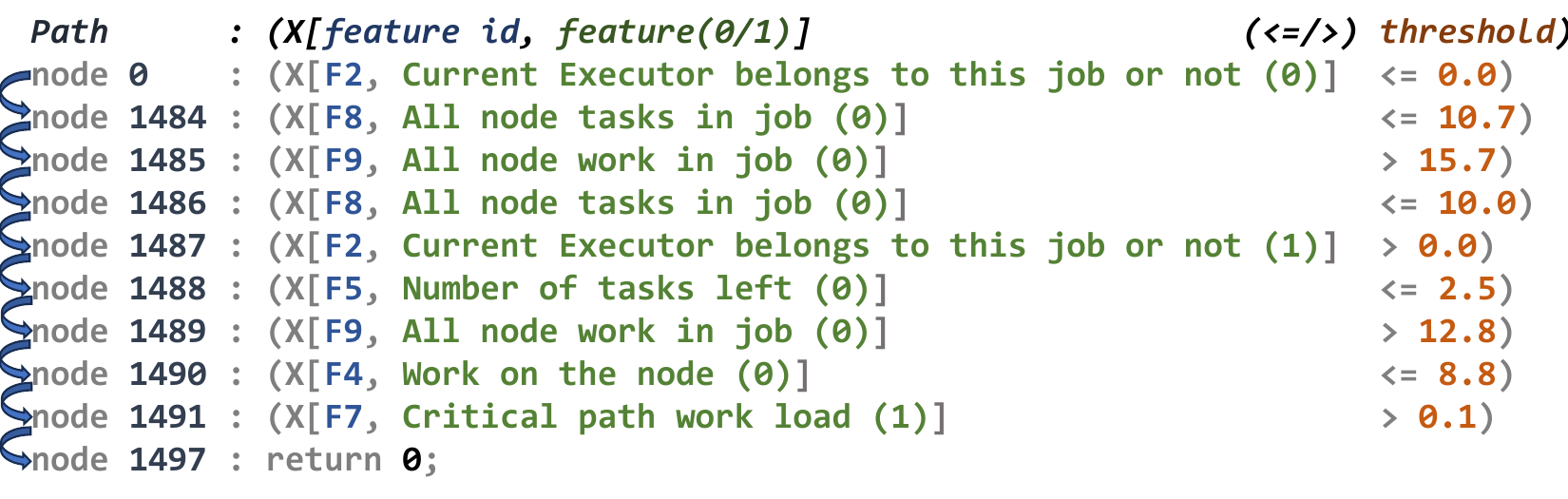}}
     \caption{TPC-H jobs}
  \end{subfigure}
  \hfill
   \begin{subfigure}[b]{0.49\linewidth}
     \centering 
     {\includegraphics[width=\linewidth]{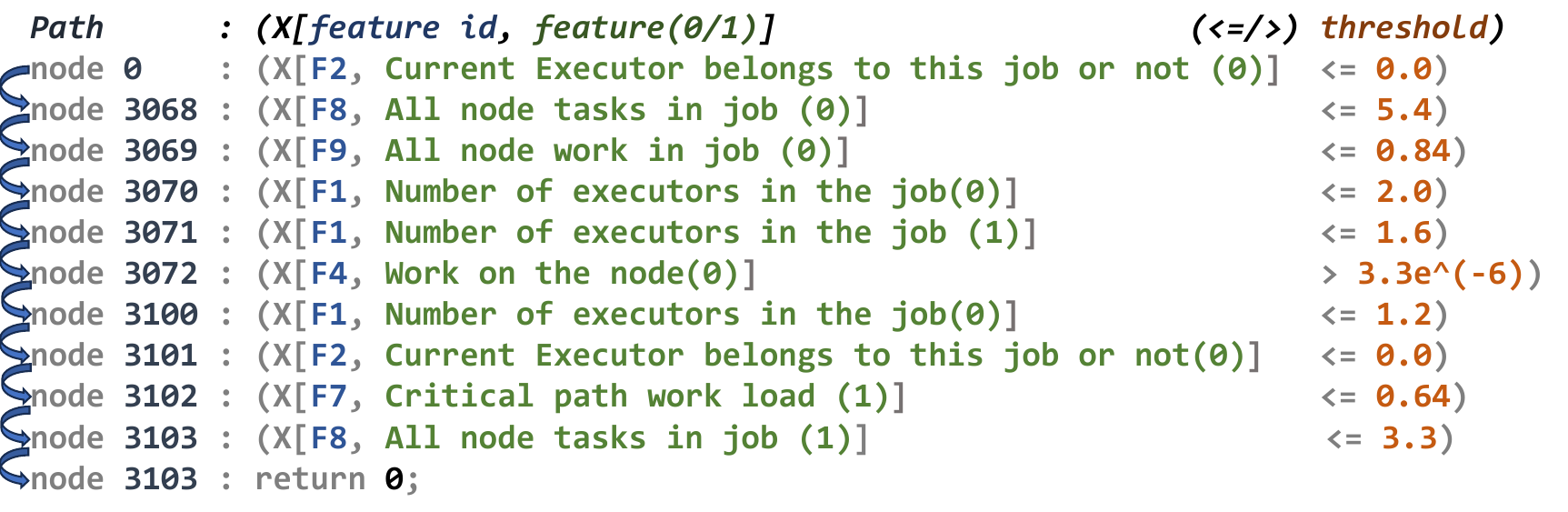}}
     \caption{Alibaba jobs}
    \end{subfigure}
  \caption{Most frequent decision tree paths.}
  \label{fig:Decision_tree_path}
\end{figure}

\noindent\textbf{Job decision path distribution in decision trees.} We show the decision trajectory distribution in Fig.~\ref{fig:Leave_distribution} as a use-case about how decision trees assist users' understanding of the scheduling decisions. Even though the trees themselves can be complex, the trajectory distribution shows certain heavy-tailed characteristics of decision paths. Fig.~\ref{fig:Decision_tree_path} 
gives the details of the most frequently traversed paths for TPC-H jobs and Alibaba jobs respectively, from which one can 
infer which job node features 
the scheduler prioritizes. For example, while the scheduler tends to prioritize workload than executor features for most TPC-H jobs, it prioritizes executor over workload features for most Alibaba jobs. This level of explanation provides meaningful information for diagnosing problems and understanding workload patterns.   

\subsection{Overall Scheduling Performance}

To evaluate scheduling performance, we consider batched arrivals, where multiple jobs start simultaneously and run to completion, and continuous arrivals, where jobs arrive based on stochastic interarrival distributions or a specific trace.

\noindent\textbf{Batched arrivals.} For TPC-H, we randomly sample 20 jobs from seven different data sizes (2, 5, 10, 20, 50, 80, and 100GB) and 22 different DAG structures, and run them on a Spark cluster with 20 executors; for Alibaba production trace, we sample 10 jobs and run them on the cluster with 50 executors. Both TPC-H and Alibaba jobs arrive in a batch, and we measure their average JCT. 

\begin{table}[h!]
\caption{Job Completion Time (JCT) for different schedulers (in s)}
\label{tab:schedulers}
\centering
\footnotesize
\begin{tabular}{|l|ccccc|}
\hline
\textbf{Scheduler} & FIFO & Fair & Decima & Decision Tree \(|g|=2\) & Decision Tree \(|g|=3\)\\
\hline
\textbf{JCT (s)} & 363.11 $\pm$ 81.08 & 263.71 $\pm$ 67.86 & 198.89 $\pm$ 46.10 & 191.87 $\pm$ 44.57 & 191.48 $\pm$ 44.71\\
\hline
\end{tabular}
\end{table}

Table~\ref{tab:schedulers} shows the executor allocation along time and job completion for TPC-H, which compares the following schedulers: (\textcolor{red}{a}) Spark's default FIFO scheduling; (\textcolor{red}{b}) The fair
scheduler that dynamically partitions executors equally between jobs; (\textcolor{red}{c}) Decima; (\textcolor{red}{d}) 
The decision tree scheduler with $|g| = 2$; 
and (\textcolor{red}{e}) The decision tree scheduler with $|g| = 3$. 
Results for the Alibaba trace are provided in Appendix~\ref{sec:appendix_batched_alibaba}.

As shown in the table, decision tree schedulers outperform non-decision tree ones. The decision tree scheduler with $|g|=3$ improves the average JCT by 47\% compared to FIFO, 27\% compared to the fair scheduler, and 4\% compared to Decima. The group size impacts performance, with a larger group size enhancing generalizability and reducing JCT. Additionally, the executor allocation strategy significantly influences the decision tree scheduler's performance. Replacing the fair allocation with Decima's parallelism limit selection significantly increases the average JCT. Our method combines Decima's node selection strategy with a heuristic-based executor allocation strategy and achieves better performance.

We further show the results of the average JCT breakdown with the job sizes measured by TPC-H data sizes. In addition, we train another decision tree model on a trace that contains all job DAG structures to evaluate the generalizability of our method. 
\begin{wrapfigure}{l}{0.5\linewidth}
    \centering
    \includegraphics[width=0.9\linewidth]{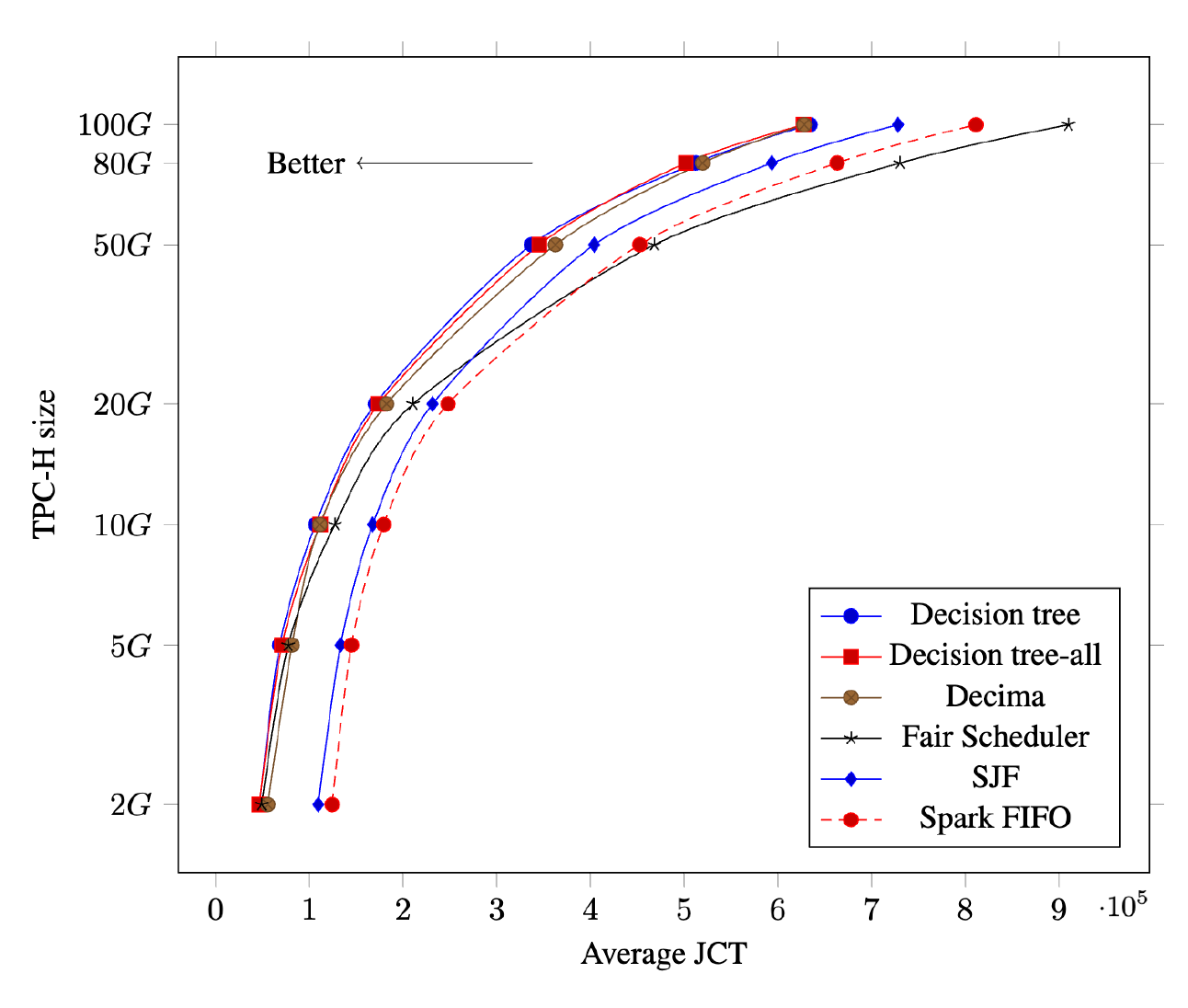}
\caption{Batched arrivals results on TPC-H: |g| = 2; decision tree-all uses all DAG types for training.}
\label{fig:Batched arrivals results}
\end{wrapfigure}
Fig.~\ref{fig:Batched arrivals results} shows the average JCT of different models over 20 experiments. We observe the following: 1) The JCT of the decision tree scheduler is slightly lower than that of the fair scheduler when the sizes of TPC-H jobs are small. The decision tree, however, shows increasing advantage when the job size increases. It outperforms the fair scheduler by 131\% with large job size. The decision tree scheduler prioritizes small jobs, the early completion of small jobs makes room for large jobs in the batch arrival case and leads to a significant performance gain measured by JCT.
2) The decision tree scheduler outperforms Decima on small jobs. On the '2G' and '5G' TPC-H queries, the decision tree scheduler outperforms Decima by 116\%. 
The advantage is mainly due to that the decision tree scheduler incorporates the fair executor allocation strategy, which is superior to the parallelism limit prediction strategy in Decima. 
3) 
There is no much obvious performance gap between the decision tree trained on the whole job DAG types and the decision tree trained on partial job DAG types. They have 
similar job JCTs on different query sizes. This shows a good generalizability of our decision tree method and indicates that a small amount of trace data can train a decision tree with good performance.

\noindent\textbf{Continuous arrivals.} We also evaluate the performance of the decision tree scheduler on streaming jobs with continuous arrivals. The result is in Appendix~\ref{sec:appendix_continuous} 

\subsection{Adapting to Edge Cases}
Finally, we examine how well our decision tree scheduler adapts to edge cases compared to Decima. We train Decima with 50 executors and 200 streaming jobs following a 25 second Poisson job arrival interval. Using two Decima models, one performing better than the other, we train the decision tree scheduler ($DT_{\text{base}}$) with the logs of the less-performing Decima model to identify edge cases. Both Decima (the better performing model) and $DT_{\text{base}}$ are then fine-tuned with edge cases over 100,000 epochs, resulting in the fine-tuned schedulers $Decima^{}_{\text{ft}}$ and $DT^{}_{\text{ft}}$ respectively. Note $DT^{*}_{\text{ft}}$ is obtained by fine-tuning ($DT_{\text{base}}$) using scheduling traces from a fair scheduler, on edge cases. The experiment considers 10 edge cases: 8 from TPC-H and 2 from the Alibaba trace, detailed in Appendix~\ref{sec:appendix_edge_cases}. Tabel~\ref{tab:edge_cases} presents the results on edge cases. It is evident that both Decima and $Decima^{*}_{\text{ft}}$ do not perform well, indicating that Decima does not handle or adapt to edge cases effectively. On the other hand, the fair scheduler performs well on all edge cases. While $DT_{\text{base}}$ does not perform well on edge cases, its fine-tuned schedulers yield better performance, comparable to or even surpassing that of the fair scheduler. The table shows the performance of fine-tuned decision tree $(DT^{*}_{\text{ft}})$ in different sizes, using $d$ to represent the maximum tree depth and $l$ the maximum number of leaves, along with the percentage improvement in average JCT. Decima has more randomness, result in a larger error bar.

\begin{table*}[h]
\caption{Fine-tuned decision tree performance on edge cases (average JCT in seconds, 50 times)}
\label{tab:edge_cases}
\centering
\resizebox{\linewidth}{!} {
\begin{tabular}{@{}|l|l|l|l|c|c|c|c|c|c|c|c|c|@{}}
\specialrule{1pt}{0pt}{0pt}
Edge cases & $Decima$ & $Decima^{*}_{\text{ft}}$ & Fair scheduler & $DT_{\text{base}}$ (d: 30; l: 1000) & $DT^{*}_{\text{ft}}$ (d: 5; l: 50) & $DT^{*}_{\text{ft}}$ (d: 10; l: 100) & $DT^{*}_{\text{ft}}$ (d: 20; l: 500) & $DT^{*}_{\text{ft}}$ (d: 30; l: 1000) & JCT $\downarrow$ \\ 
    \specialrule{1pt}{0pt}{0pt}
    1 (Appendix fig \ref{fig:case_1}) & 21.58 $\pm$ 3.45 & 21.62 $\pm$ 2.32 & 16.58 $\pm$ 0.32& \textcolor{blue}{\textbf{18.03 $\pm$ 0.56}} & 16.52 $\pm$ 0.24 & \textcolor{red}{\textbf{16.20 $\pm$ 0.22}} & 16.61 $\pm$ 0.39 & 16.65 $\pm$ 0.30 & 10.14\% \\
    \cline{1-10}
    2 (Appendix fig \ref{fig:case_2}) & 22.33 $\pm$ 2.41 & 21.62 $\pm$ 3.55 & 17.66 $\pm$ 0.28 & \textcolor{blue}{\textbf{20.59 $\pm$ 0.85}} & \textcolor{red}{\textbf{17.64 $\pm$ 0.36}} & 17.75 $\pm$ 0.32 & 17.97 $\pm$ 0.35 & 17.90 $\pm$ 0.38 & 14.34\% \\
    \cline{1-10}
    3 (Appendix fig \ref{fig:case_3}) & 28.36 $\pm$ 0.41 & 28.44 $\pm$ 2.34 & 19.79 $\pm$ 0.27 & \textcolor{blue}{\textbf{20.86 $\pm$ 0.39}} & 20.12 $\pm$ 0.32 & \textcolor{red}{\textbf{19.94 $\pm$ 0.3}} & 21.38 $\pm$ 0.36 & 21.26 $\pm$ 0.32 & 4.42\% \\
    \cline{1-10}
    4 (Appendix fig \ref{fig:case_4}) & 19.68 $\pm$ 2.03 & 18.88 $\pm$ 3.33 & 15.55 $\pm$ 0.39 & \textcolor{blue}{\textbf{20.04 $\pm$ 0.72}} & 17.74 $\pm$ 0.71 & 15.81 $\pm$ 0.48 & \textcolor{red}{\textbf{15.54 $\pm$ 0.45}} & 15.54 $\pm$ 0.44 & 22.49\% \\
    \cline{1-10}
    5 (Appendix fig \ref{fig:case_5}) & 18.89 $\pm$ 1.76 & 18.72 $\pm$ 2.27 & 15.88 $\pm$ 0.33 & \textcolor{blue}{\textbf{17.70 $\pm$ 0.48}} & 16.54 $\pm$ 0.57 & 15.98 $\pm$ 0.38 & 15.96 $\pm$ 0.56 & \textcolor{red}{\textbf{15.90 $\pm$ 0.36}} & 10.17\% \\
    \cline{1-10}
    6 (Appendix fig \ref{fig:case_6}) & 20.03 $\pm$ 2.48 & 19.53 $\pm$ 0.39 & 15.43 $\pm$ 0.26 & \textcolor{blue}{\textbf{17.29 $\pm$ 0.41}} & 15.77 $\pm$ 0.28 & \textcolor{red}{\textbf{15.47 $\pm$ 0.32}} & 15.51 $\pm$ 0.24 & 15.58 $\pm$ 0.23 & 10.50\% \\
    \cline{1-10}
    7 (Appendix fig \ref{fig:case_7}) & 38.93 $\pm$ 3.65 & 39.92 $\pm$ 2.09 & 35.60 $\pm$ 0.62 & \textcolor{blue}{\textbf{39.43 $\pm$ 1.03}} & 35.61 $\pm$ 0.35 & 35.48 $\pm$ 0.66 & \textcolor{red}{\textbf{35.46 $\pm$ 0.72}} & 35.67 $\pm$ 0.72 & 10.07\% \\
    \cline{1-10}
    8 (Appendix fig \ref{fig:case_8}) & 18.63 $\pm$ 0.36 & 18.25 $\pm$ 3.41 & 15.13 $\pm$ 0.36 & \textcolor{blue}{\textbf{17.52 $\pm$ 0.56}} & 15.29 $\pm$ 0.55 & 15.79 $\pm$ 0.35 & \textcolor{red}{\textbf{15.15 $\pm$ 0.45}} & 15.17 $\pm$ 0.46 & 13.57\% \\
    \cline{1-10}
    9 (Appendix fig \ref{fig:case_9}) & 17.70 $\pm$ 1.76 & 17.22 $\pm$ 1.15 & 15.89 $\pm$ 0.43 & \textcolor{blue}{\textbf{16.40 $\pm$ 0.25}} & 16.25 $\pm$ 0.65 & 15.89 $\pm$ 0.42 & 15.90 $\pm$ 0.26 & \textcolor{red}{\textbf{15.89 $\pm$ 0.31}} & 3.09\% \\
    \cline{1-10}
    10 (Appendix fig \ref{fig:case_10}) & 3.77 $\pm$ 0.23 & 3.76 $\pm$ 0.30 & 3.39 $\pm$ 0.18 & \textcolor{blue}{\textbf{3.41 $\pm$ 0.12}} & 3.401 $\pm$ 0.28 & 3.40 $\pm$ 0.22 & \textcolor{red}{\textbf{3.38 $\pm$ 0.27}} & 3.39 $\pm$ 0.12 & 0.878\% \\
    
\specialrule{1pt}{0pt}{0pt}
\end{tabular}
}
\end{table*}

For batched arrivals, the decision tree scheduler fine-tuned with edge cases achieves nearly performance with an average JCT of 194202.30 ms as the non-fine-tuned version with an average JCT of 191871.99 ms of Decision tree |g|=2 for the overall workload, indicating the scheduler's adaptability and generalizabilty. For continuous arrivals, the fine-tuned decision tree scheduler (|g|=2) achieves an average JCT of 56779.28ms, better than the non-fine-tuned scheduler's 57670.62ms (Fig. \ref{fig:Scheduling_trace}).

\section{Conclusion} 
As deep learning is utilised increasingly in system research, it is worth noting the significant gap between the strategy learned by a DNN model and its deployment in practice. Specifically, the DNN model predictions are difficult to interpret and hard to patch when problems occur. In this paper, we proposed a method to distill interpretable models from a complex DNN model to address the challenge in developing optimised scheduling algorithm in data processing clusters. More importantly, we gave a method to efficiently improve the scheduler's performance on edge cases through interpretable model tuning. Our decision tree scheduler 
outperforms the state-of-the-art DNN-based scheduler, esp. with large jobs. 

\paragraph{Limitations} Even though a decision tree is easy to tune to adapt to edge cases, we rely on heuristic scheduling strategies to produce a reference better schedule for edge cases. This may limit the tuning performance.

\bibliographystyle{plainnat} 
\bibliography{explainsys}

\newpage
\appendix

\section{Appendix / supplemental material}



\subsection{Dataset Generation for Interpretability Study}
\label{sec:appendix_interpretability_dataset_details}

For TPC-H, we randomly sample jobs from seven different data sizes (2, 5, 10, 20, 50, 80, and 100GB) and all 22 TPC-H queries or DAG structures, creating two non-overlapping partitions of 500 jobs each. We run Decima on these partitions, and use the logs to obtain training and test sets for the decision tree scheduler. Both the training and test sets consist of the combined features of node comparators and the selected nodes for each scheduling event. By setting the group size, $|g|$ to 2, we have 252,472 training pairs. Each set includes the combined features of node comparators and selected nodes for each scheduling event. With a group size $|g|$ of 2, we obtain 252,472 training pairs. With $|g|$ set to 3, we get 1,796,180 training triplets. The test set contains 119,414 pairs and 456,420 triplets. 

\subsection{Batched Arrivals: Alibaba Trace}
\label{sec:appendix_batched_alibaba}

\begin{figure*}[htbp]
    \centering
    \includegraphics[width=\linewidth]{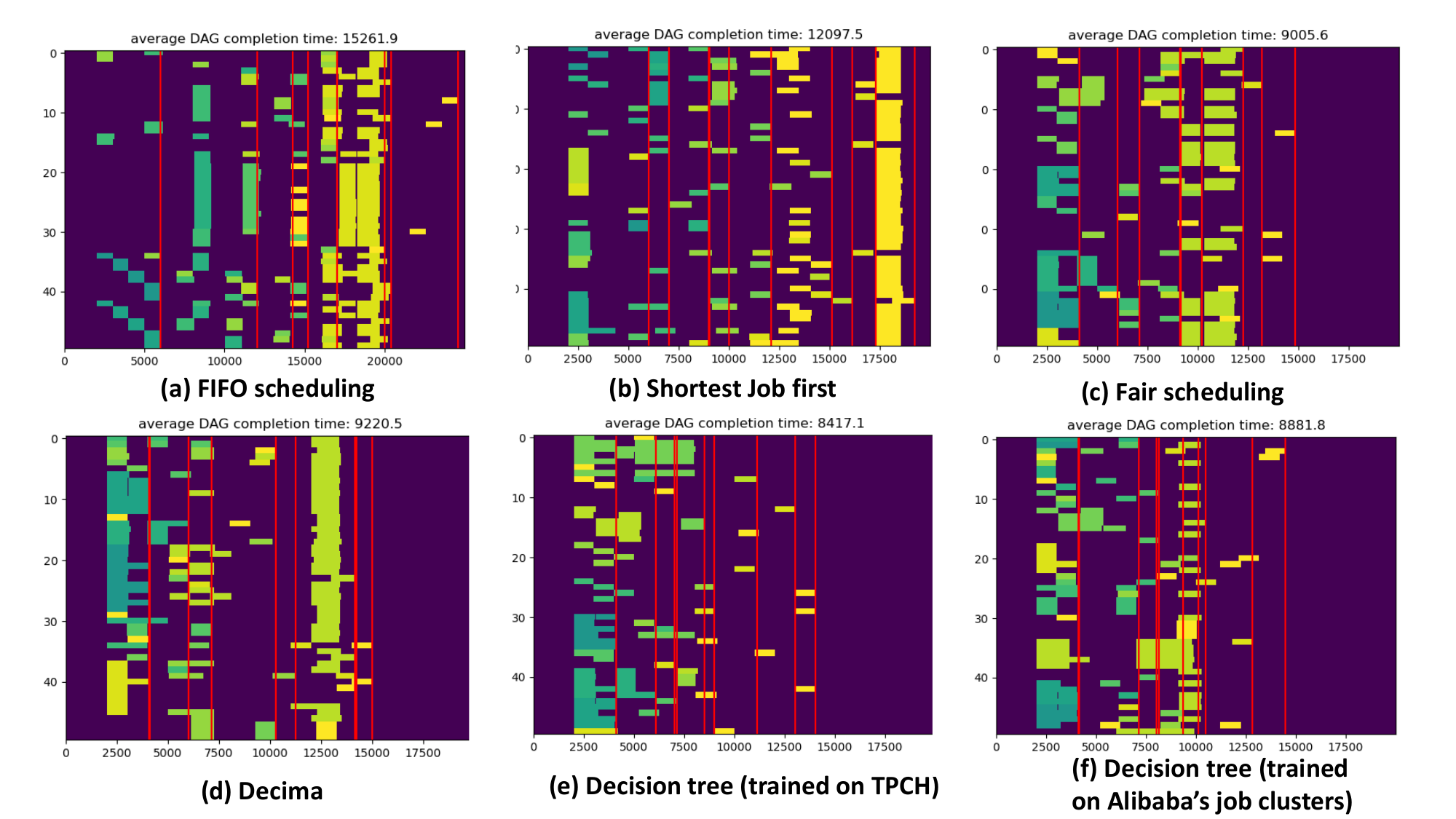}
    \caption{Model generalisability results (job arrival starts from time 2500): the Decision tree scheduler trained on TPC-H improves average JCT of 10 random Alibaba production jobs by 44.8 $\%$ over Spark’s FIFO scheduler, by 30.4 $\%$ over shortest job first, by 6.5 $\%$ over a fair scheduler, and 8.7 $\%$ over Decima on a cluster with
    50 task slots (executors). Different queries in different colors; vertical red lines are job completions; purple means idle. Different to the experiment on TPC-H jobs, we additionally visualize scheduling details under Shortest job first (SJF) for comparing more algorithms on a real job trace.}
\label{fig:Scheduling_details_Alibaba}
\end{figure*}

We apply the model trained on TPC-H on the Alibaba production job trace to evaluate the generalizability of our decision tree model. Fig.~\ref{fig:Scheduling_details_Alibaba} shows that our decision tree scheduler consistently outperforms the baselines. In Fig.~\ref{fig:Scheduling_details_Alibaba} (e) and (f), we compare the performance difference between a model trained on TPC-H and a model trained on Alibaba production trace, using test data containing workload from the Alibaba trace only.  
The model trained on TPC-H slightly outperforms the one trained on Alibaba trace. It is mainly because the TPC-H jobs are more diverse, which enables the model to select optimised decision paths for jobs in the test data. 
The results show the good generalizability of our decision tree scheduler on unseen job patterns.

\subsection{Continuous Arrivals}
\label{sec:appendix_continuous}

\begin{figure*}[tb]
    \centering
    \includegraphics[width=\linewidth]{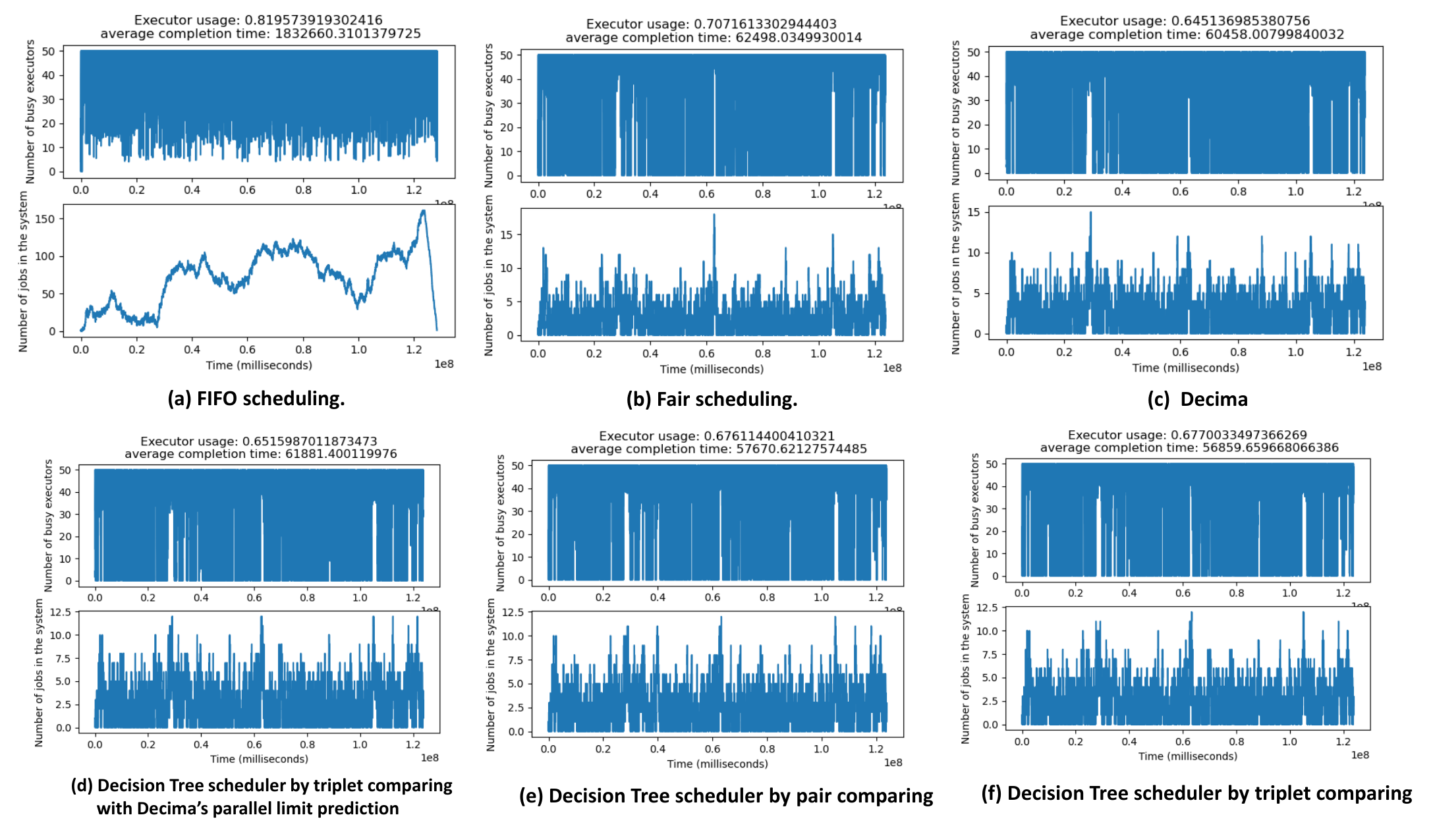}
    \caption{Decision tree scheduler performance comparison on TPC-H streaming jobs: executor number -- 50. The triplet decision tree model improves average JCT of 5000 streaming TPC-H queries by 97$\%$ over Spark’s FIFO scheduler, by 9$\%$ over the Fair scheduler, by 6$\%$ over Decima. 
    The maximum executor utilization is 50. The number of concurrent jobs in the system is collected at each time point.
}
\label{fig:Scheduling_trace}
\end{figure*}

\noindent\textbf{Continuous arrivals.} We also evaluate the performance of the decision tree scheduler on streaming jobs with continuous arrivals. We 
randomly sample 5,000 TPC-H jobs from seven 
different sizes and 5000 Alibaba’s production jobs which are unseen in training jobs. We generate their arrival intervals using a Poisson process with a mean of 25 seconds for TPC-H jobs and 2.5 seconds for Alibaba’s production jobs (as the Alibaba jobs' sizes are smaller). 

Figure \ref{fig:Scheduling_trace} shows that the decision tree scheduler outperforms heuristic-based schedulers and Decima on TPC-H jobs. It maintains a lower maximum number of concurrent jobs in the system and a lower variance in the number of concurrent jobs at each time point. The average maximum number of concurrent jobs in the system at each time under the fair scheduler and Decima is 19 and 15 respectively while the decision tree scheduler has only 12.5 concurrent jobs in the system in average at each time point. This is an indication of its efficient resource use.

Similar to Decima, the performance gain of the decision tree schedulers also comes from early completion of smaller jobs. But unlike Decima, our decision tree schedulers allocate executors fairly among all concurrent jobs rather than allocate extra executors to the small jobs as Decima does. It is clear that Decima fails to learn an optimal executor allocation strategy, which indicates the limitation of a purely data-driven approach for learning scheduling policies. Our decision tree scheduler instead allows the injection of heuristic strategies to an interpretable model for further performance improvement.
We achieve this by distilling node selection strategy from the DNN model in Decima and replacing its executor allocation strategy 
with fair scheduling. 
To demonstrate the difference, we downgrade our executor allocation strategy in the triplet decision tree ($|g|=3$) to the one used by Decima. Fig.~\ref{fig:Scheduling_trace}(d) shows the resulting JCT increases to a level close to the one produced by the original Decima scheduler in Fig.~\ref{fig:Scheduling_trace}(c), 
in contrast to the significant lower JCT achieved in Fig.~\ref{fig:Scheduling_trace}(e) and (f). This clearly shows the importance to develop an interpretable model to enable the use of heuristics or domain knowledge for improvement and problem diagnostics.

\begin{figure*}[tb]
    \centering
    \includegraphics[width=0.95\linewidth]{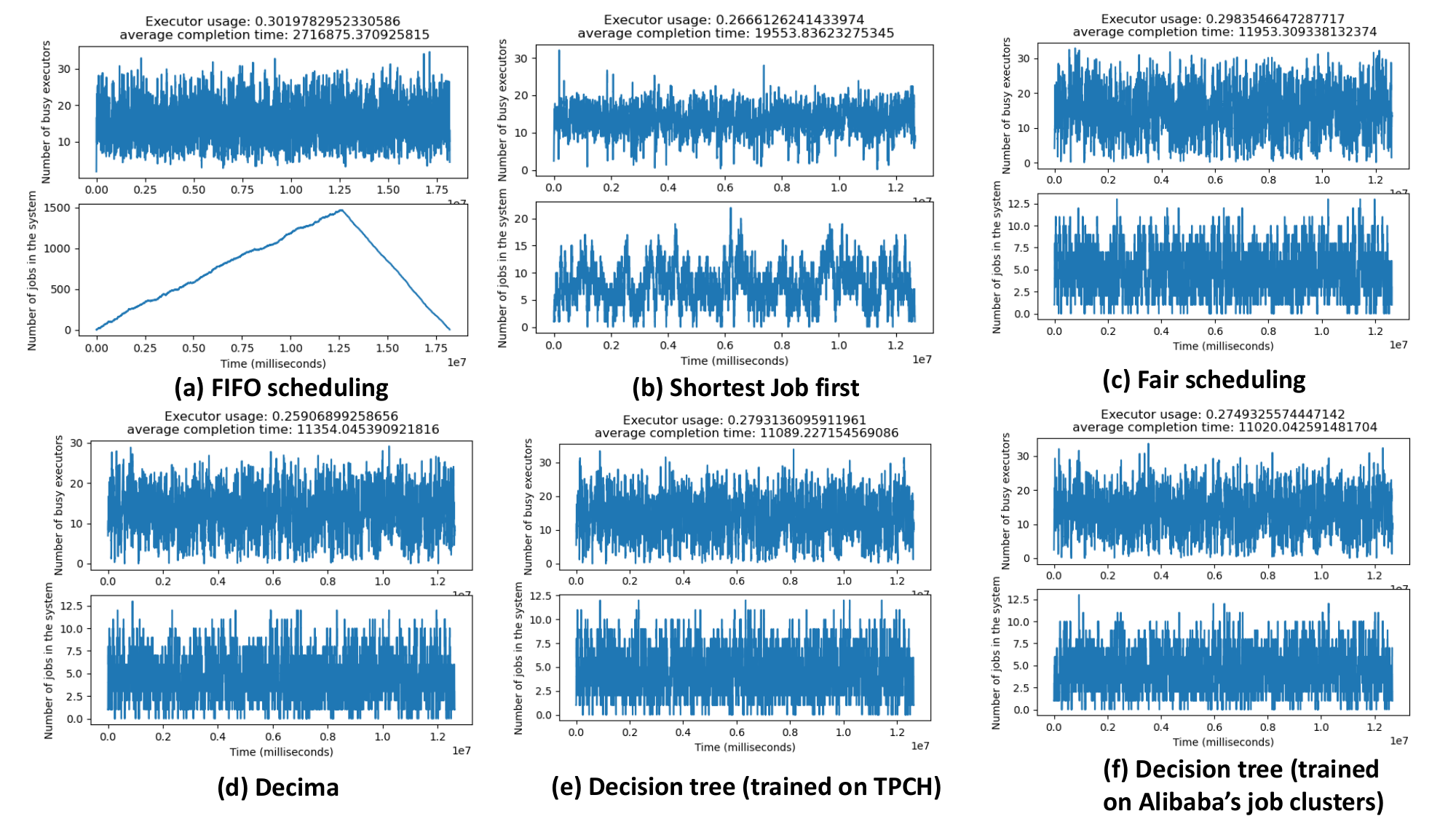}
    \caption{Model generalisability results for streaming jobs on Alibaba jobs: executor number -- 50. the decision tree model trained on Alibaba jobs improves average JCT of 5000 streaming Alibaba queries by 99 $\%$ over Spark’s FIFO scheduler, by 43 $\%$ over the SJF (Shortest Job First), by 7.8 $\%$ over the fair scheduler, by 2.8 $\%$ over Decima. 
    The model trained on TPC-H trace slightly outperforms the one trained on Alibaba trace, which indicates the good generalizability of our decision tree scheduler on unseen job patterns.
}
\label{fig:Alibaba_scheduling_trace}
\end{figure*}

On Alibaba’s production workload, we test the Decision tree scheduler obtained by two different ways. Fig.~\ref{fig:Alibaba_scheduling_trace}(e) is the decision tree scheduler trained from the scheduling trace of Decima on TPC-H jobs and Fig.~\ref{fig:Alibaba_scheduling_trace}(f) is the decision tree scheduler trained from the scheduling trace of Decima on Alibaba's production jobs. Although they have different data 
distributions, the decision tree scheduler obtained by simulating Decima's trace on TPC-H jobs also performs well on Alibaba's production jobs. There is no obvious performance gap between these two Decision tree schedulers, which 
indicates the good generalizability of our decision tree scheduler. 

We did our experiments on M2 CPU.

\subsection{Edge Cases}
\label{sec:appendix_edge_cases}
\clearpage

\begin{figure*}[tb]
    \centering
    \includegraphics[width=\linewidth]{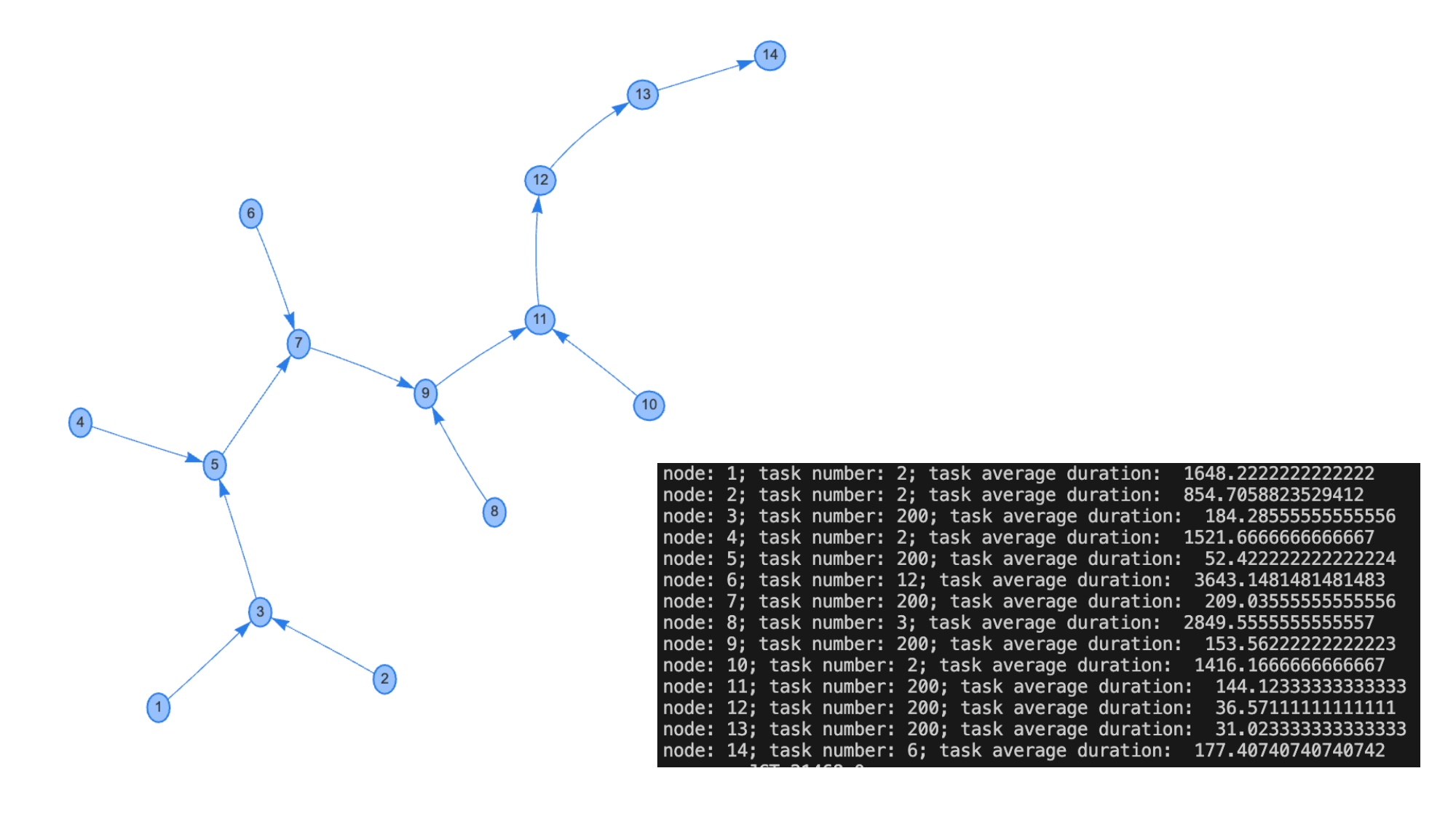}
    \caption{Case 1.
}
\label{fig:case_1}
\end{figure*}

\begin{figure*}[tb]
    \centering
    \includegraphics[width=\linewidth]{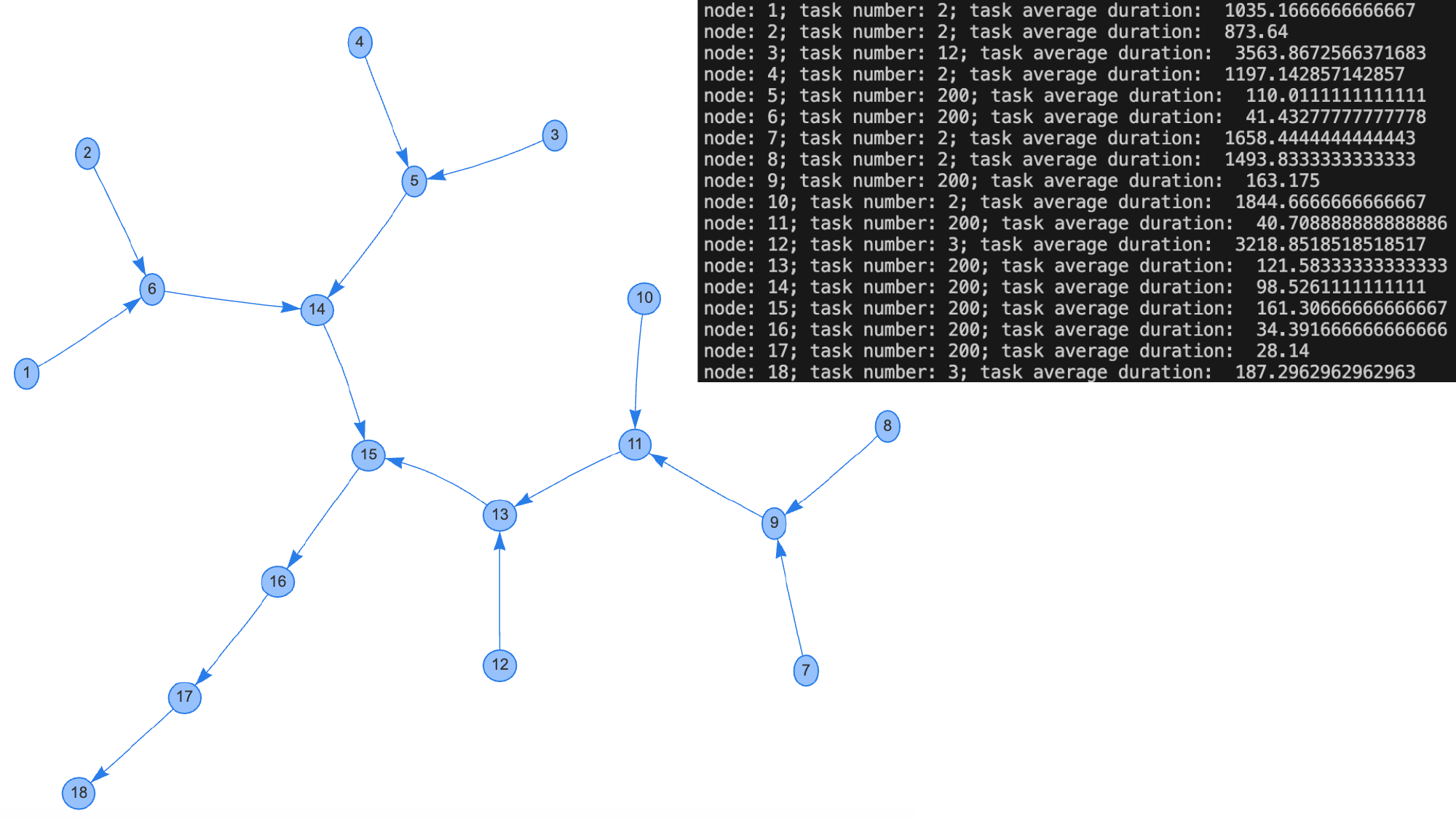}
    \caption{Case 2.
}
\label{fig:case_2}
\end{figure*}

\begin{figure*}[tb]
    \centering
    \includegraphics[width=\linewidth]{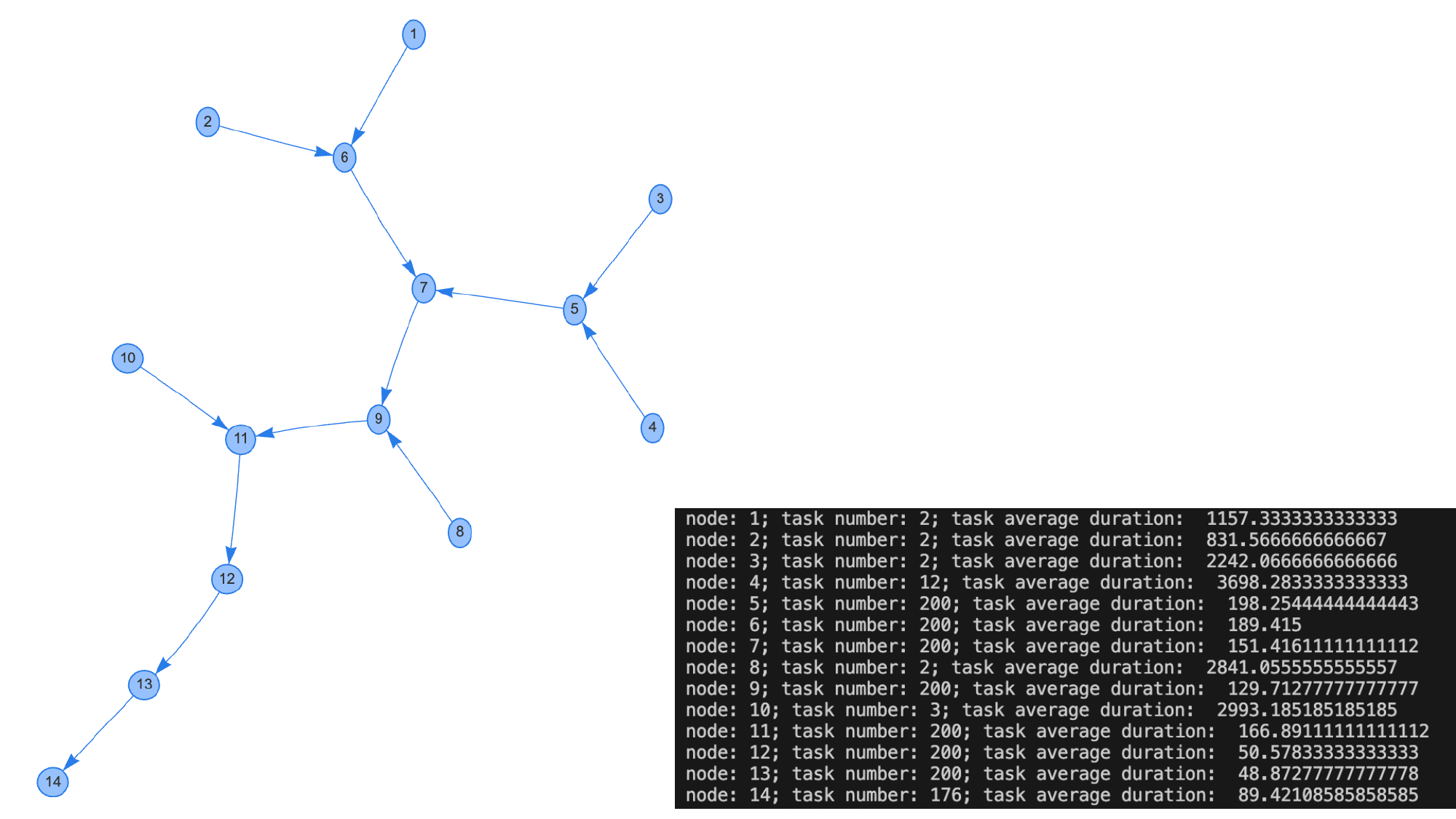}
    \caption{Case 3.
}
\label{fig:case_3}
\end{figure*}

\begin{figure*}[tb]
    \centering
    \includegraphics[width=\linewidth]{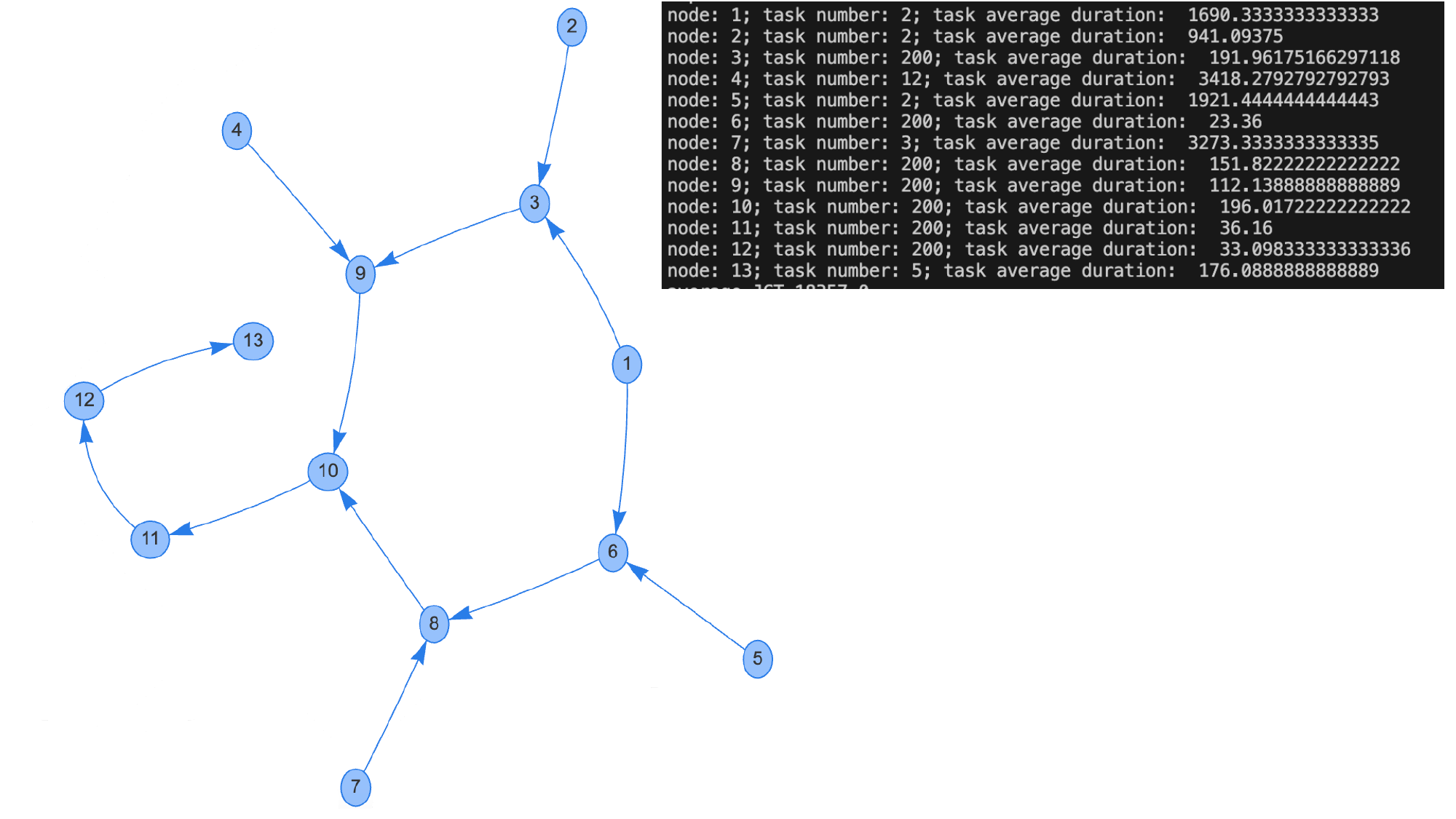}
    \caption{Case 4.
}
\label{fig:case_4}
\end{figure*}

\begin{figure*}[tb]
    \centering
    \includegraphics[width=\linewidth]{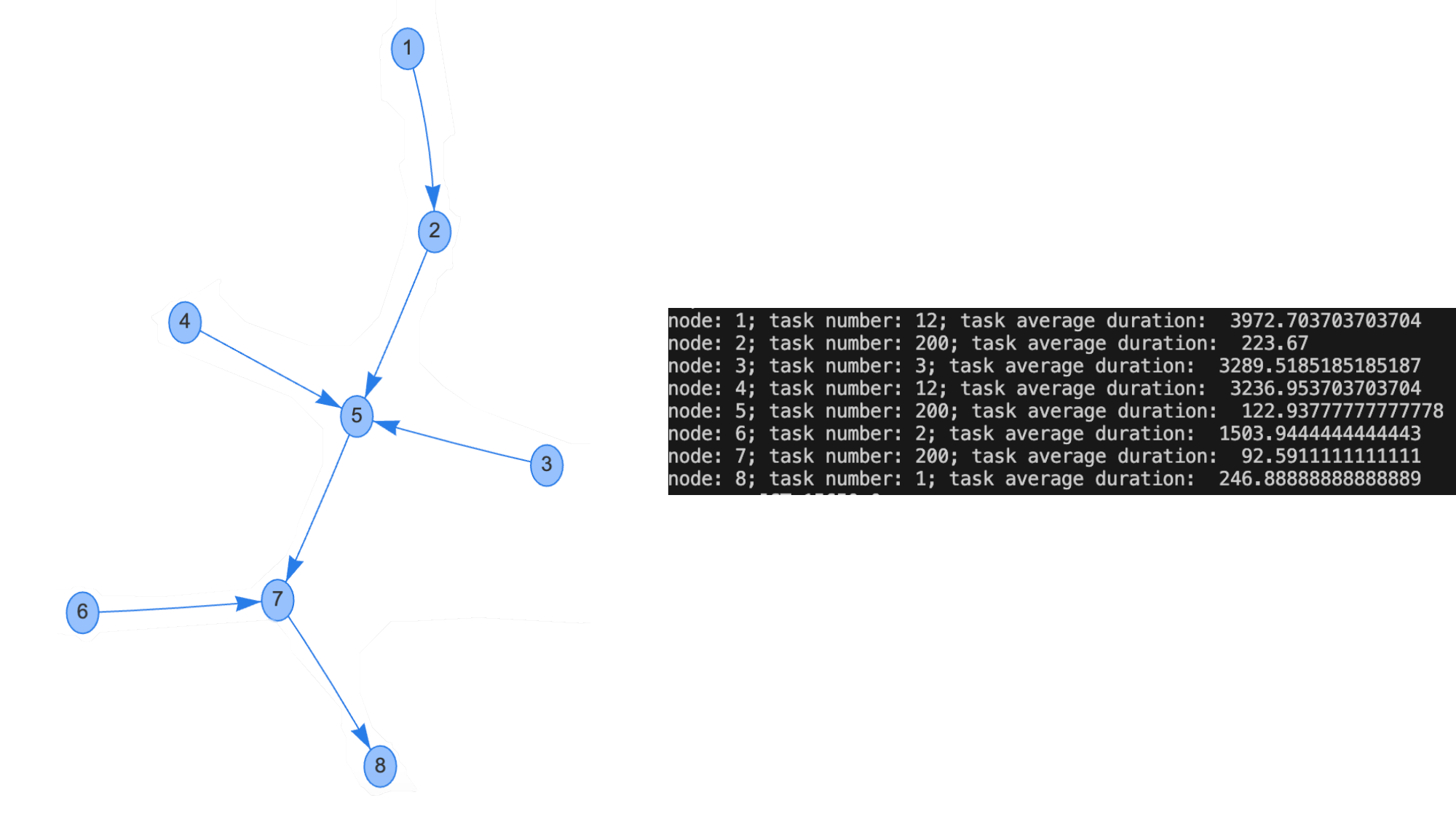}
    \caption{Case 5.
}
\label{fig:case_5}
\end{figure*}

\begin{figure*}[tb]
    \centering
    \includegraphics[width=\linewidth]{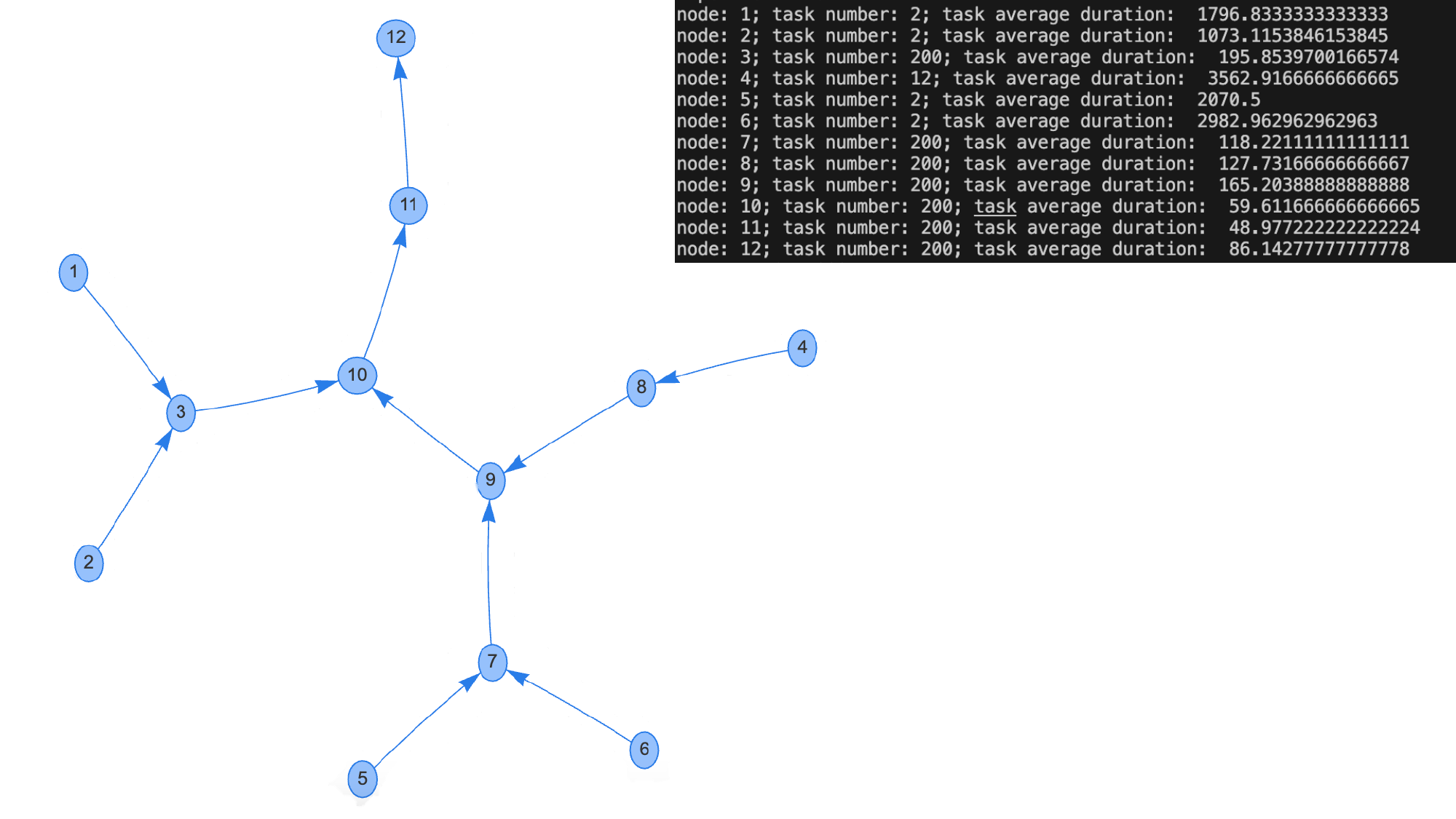}
    \caption{Case 6.
}
\label{fig:case_6}
\end{figure*}

\begin{figure*}[tb]
    \centering
    \includegraphics[width=\linewidth]{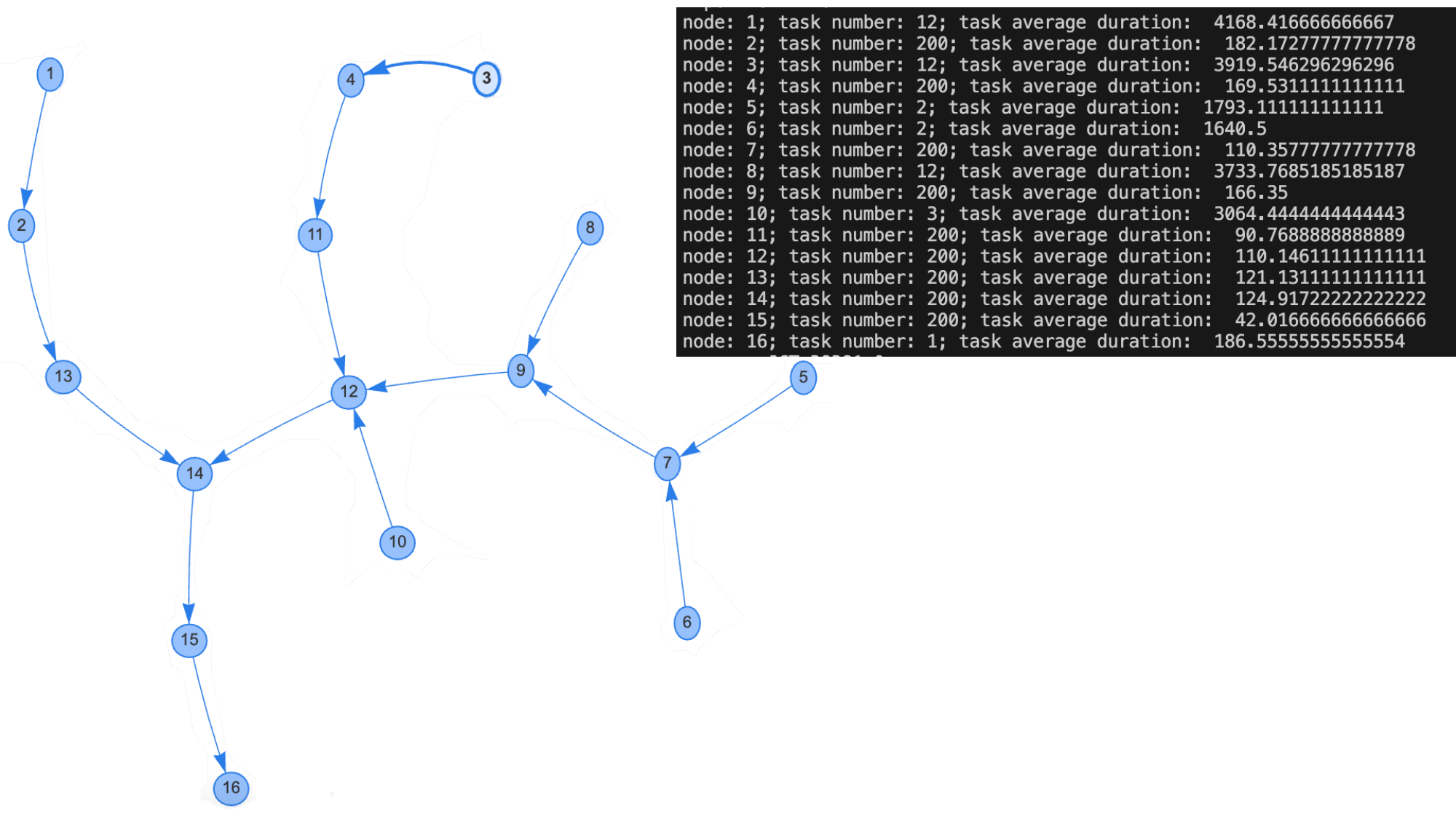}
    \caption{Case 7.
}
\label{fig:case_7}
\end{figure*}

\begin{figure*}[tb]
    \centering
    \includegraphics[width=\linewidth]{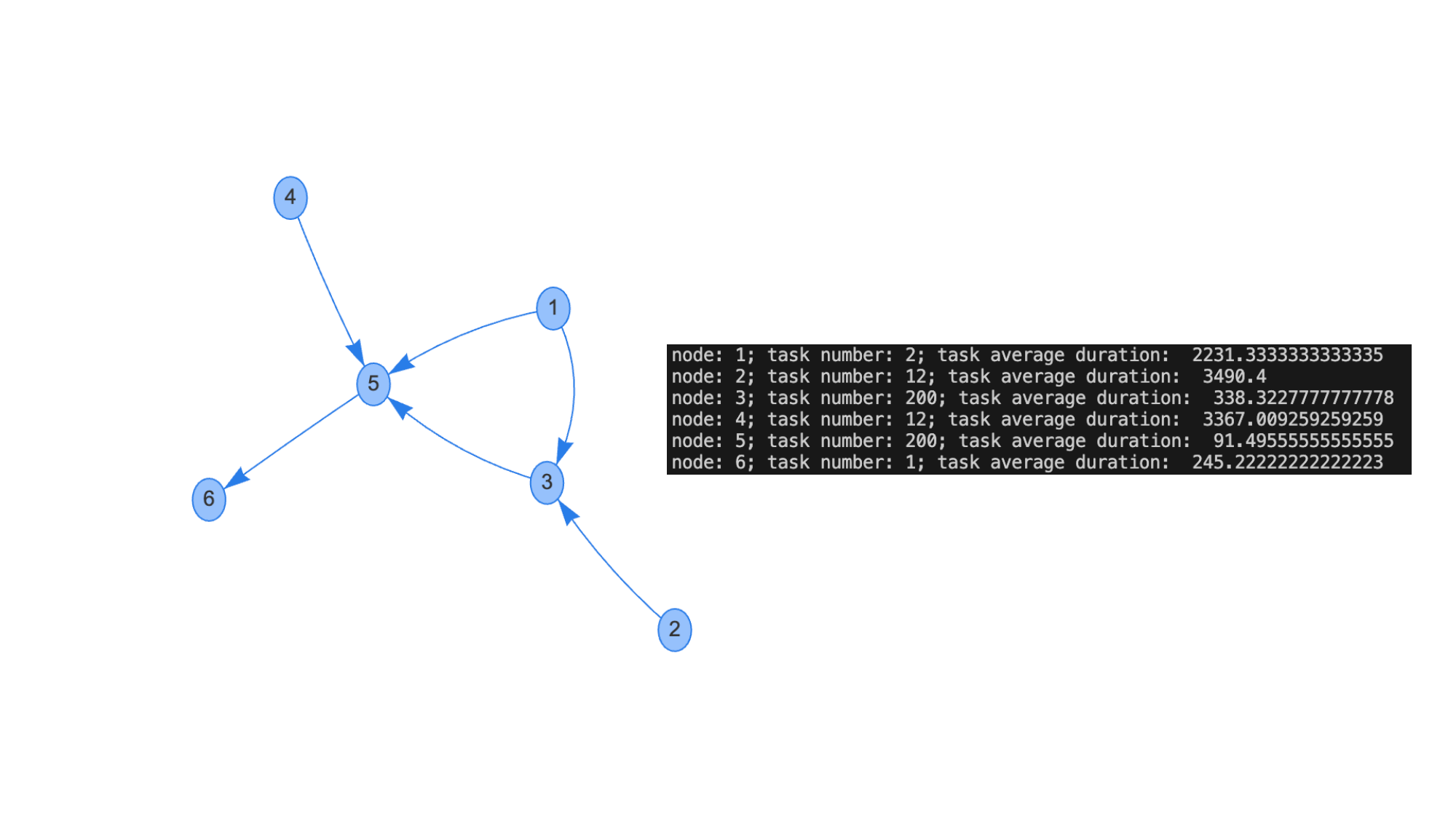}
    \caption{Case 8.
}
\label{fig:case_8}
\end{figure*}

\begin{figure*}[tb]
    \centering
    \includegraphics[width=\linewidth]{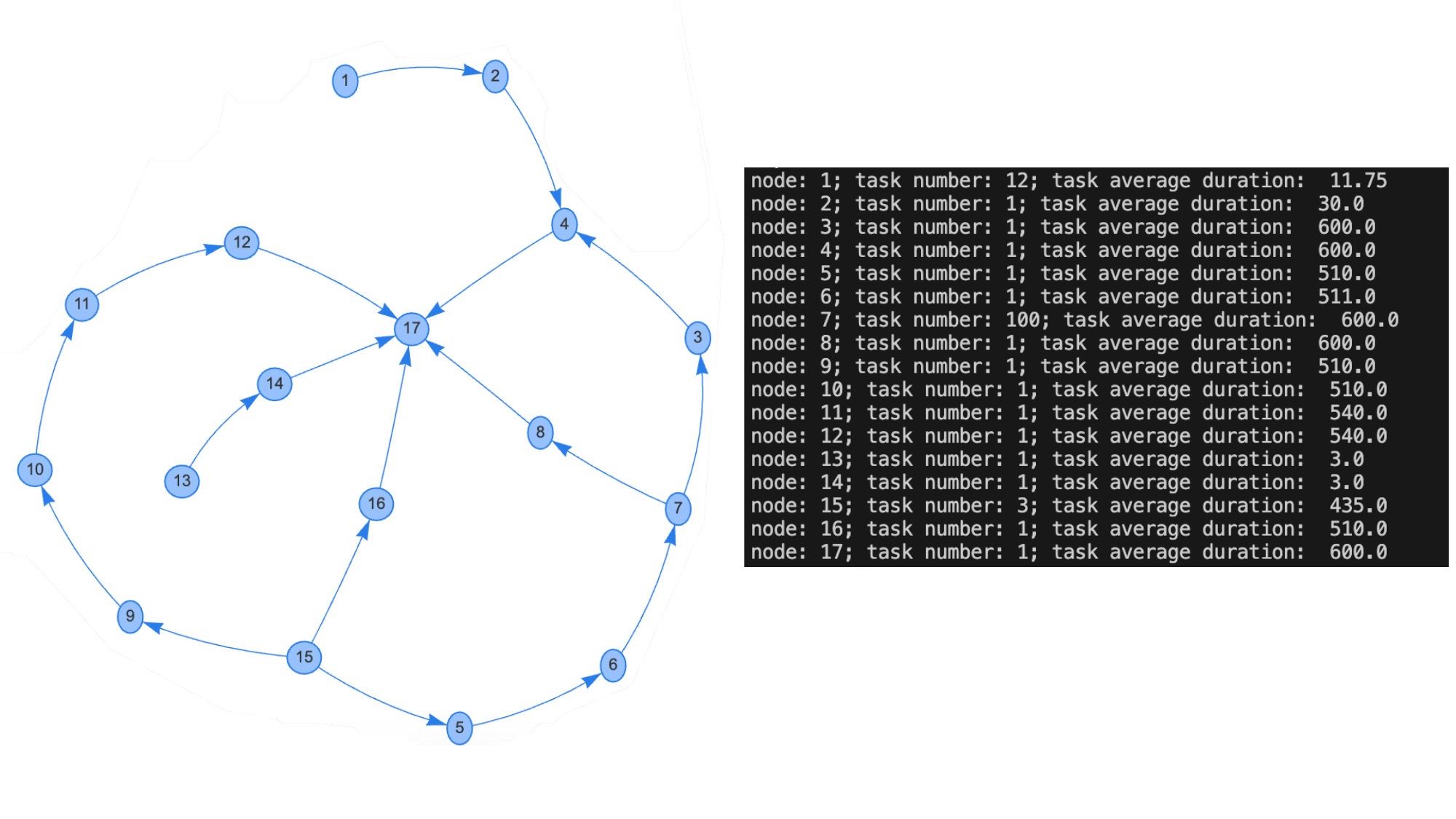}
    \caption{Case 9.
}
\label{fig:case_9}
\end{figure*}

\begin{figure*}[tb]
    \centering
    \includegraphics[width=\linewidth]{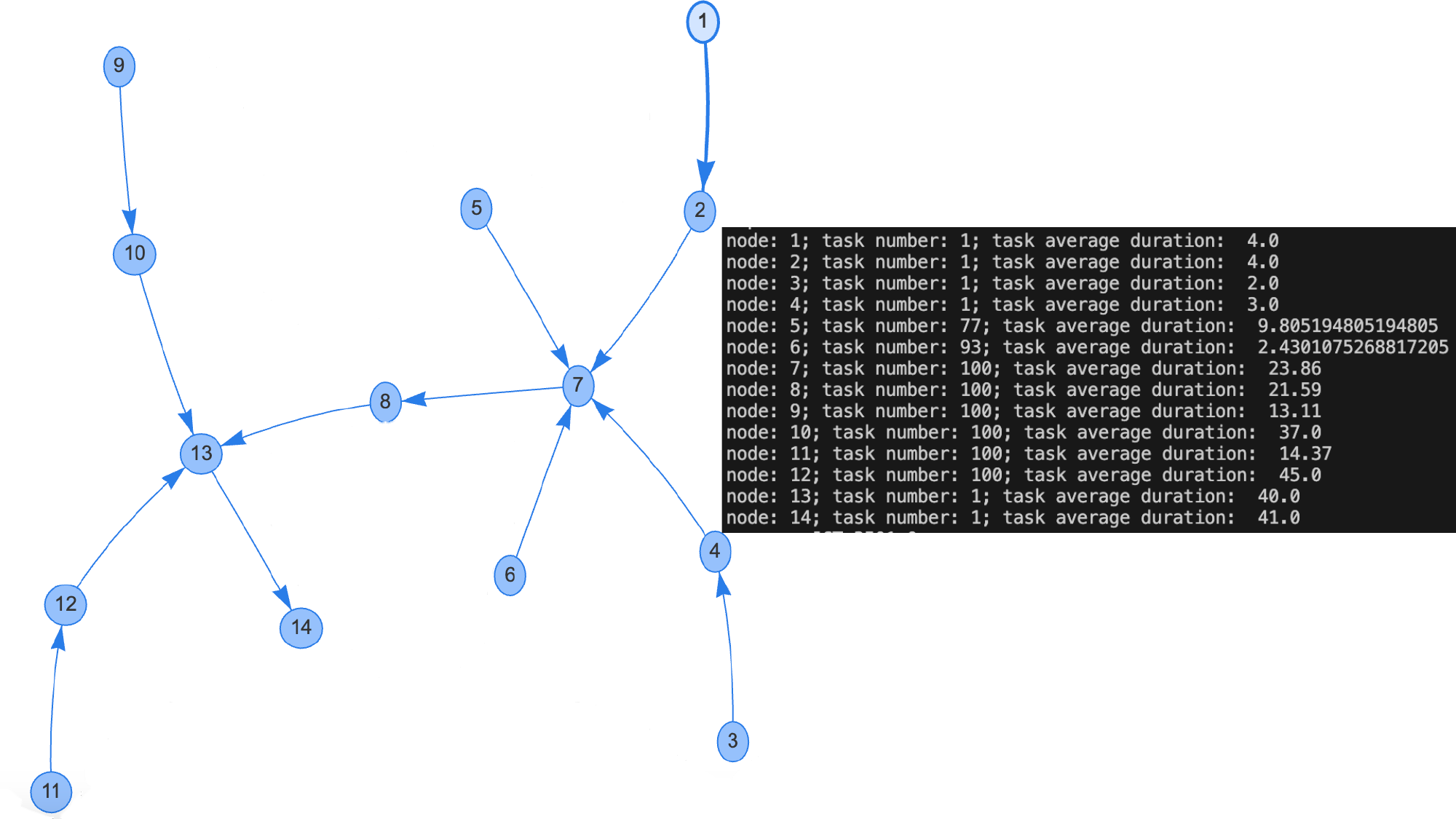}
    \caption{Case 10.
}
\label{fig:case_10}
\end{figure*}

\end{document}